\documentclass[twocolumn,iop, aps,superscriptaddress,showpacs]{revtex4-1}

\usepackage{mathptmx}
\usepackage{subfigure}
\usepackage{psfrag,graphicx}
\usepackage{dcolumn}
\usepackage{amsmath,amssymb}
\usepackage{bm}
\usepackage{color}
\usepackage{latexsym}
\usepackage{color}
\usepackage[english]{babel}
\usepackage{latexsym}

\usepackage{subfigure} 
\usepackage{amsmath} 
\usepackage{amssymb} 
\usepackage{amsfonts}
\usepackage{bm}
\usepackage{natbib}
\usepackage{epstopdf}
\usepackage{appendix}
\usepackage{changes}
\usepackage{bbold}

\definecolor{mygrey}{gray}{0.35}
\definecolor{myblue}{rgb}{0.2,0.2,0.8}
\definecolor{myzard}{cmyk}{0,0,0.05,0}
\definecolor{mywhite}{rgb}{1,1,1}
\definecolor{myred}{rgb}{1,0.,0.3}

\usepackage[colorlinks=true,citecolor=myblue,linkcolor=myred]{hyperref}

\def\be{\begin{equation}}
\def\ee{\end{equation}}
\def\ba{\begin{align}}
\def\enda{\end{align}}
\def\bi{\begin{itemize}}
\def\ei{\end{itemize}}

 \def\ee{\mathord{\rm e}}

\def\omegar { \omega_{ \mathrm{r}  } }
\def\Ir { I_{ \mathrm{r}  } }
\def\omegas { \omega_{ \mathrm{s}  } }

 \def\ee{\mathord{\rm e}}

\renewcommand{\ee}{{\rm e}}

\def\beq{\begin{equation}}
\def\beq{\begin{equation}}
\def\eeq{\end{equation}}

\newcommand{\vect}[1]{\bm #1}

\begin{document}

\title[Short Title]{Overcoming resolution limits with quantum sensing}

\author{T. Gefen}
\affiliation{Racah Institute of Physics, The Hebrew University of Jerusalem, Jerusalem 
91904, Givat Ram, Israel}
\author{A. Rotem}
\affiliation{Racah Institute of Physics, The Hebrew University of Jerusalem, Jerusalem 
91904, Givat Ram, Israel}
\author{A. Retzker}
\affiliation{Racah Institute of Physics, The Hebrew University of Jerusalem, Jerusalem 
91904, Givat Ram, Israel}
\date{\today}

\begin{abstract}
{ The field of quantum sensing explores the use of quantum phenomena to measure a broad range of physical quantities, 
of both static and time-dependent types.
 An important figure of merit for time dependent signals is the spectral resolution, i.e. the
ability to resolve two different frequencies.
Here we study this problem, and develop new superresolution methods that rely on quantum features. 
We first formulate a general criterion for superresolution in quantum problems. 
Inspired by this, we show that quantum detectors can resolve two frequencies from incoherent segments of the signal, irrespective of their separation, in contrast to what is known about classical detection schemes.
The main idea behind these methods is to overcome the vanishing distinguishability in resolution problems by nullifying the projection noise.   
}
\end{abstract}
\maketitle

\section*{Introduction}
Quantum metrology and quantum sensing \cite{giovannetti2011advances,degen2017quantum} study parameter estimation limits in various physical
systems by employing the fundamental laws of quantum physics.
In particular this field seeks to optimize precision by utilizing quantum effects that have no classical analogs (such as entanglement and squeezing \cite{bollinger1996optimal,itano1993quantum}).


A unique feature of quantum sensing is the ability to apply coherent control to the probe and vary the measurement basis.
In particular this provides the ability to nullify the measurement projection noise. However, the contribution of this phenomenon to estimation problems has received scant attention. 

In this paper we highlight this feature and show that it is a critical resource primarily for resolution problems, that can improve precision by orders of magnitude.
Resolution problems are ubiquitous and highly important in science \cite{rayleigh1879xxxi,hannan1989resolution,hell1994breaking,bettens1999model,van2002high,tsang2016quantum,nair2016far, lupo2016ultimate,barabell1998performance,glentis2014sar}, and roughly speaking are characterized by vanishing distinguishability; i.e, the sensitivity to the seperation between two close objects or frequencies vanishes as these get close enough.
This effect usually results in divergent uncertainty, leading to a resolution limit.
We show that it is possible to overcome the vanishing distinguishability by making the projection noise vanish as well, through a suitable control. These two effects can cancel each other out, leading to a finite uncertainty. We show that this is a general method to overcome resolution limits in quantum sensing.

\begin{figure}[t]
\begin{center}
\subfigure[]{\includegraphics[width=7 cm]{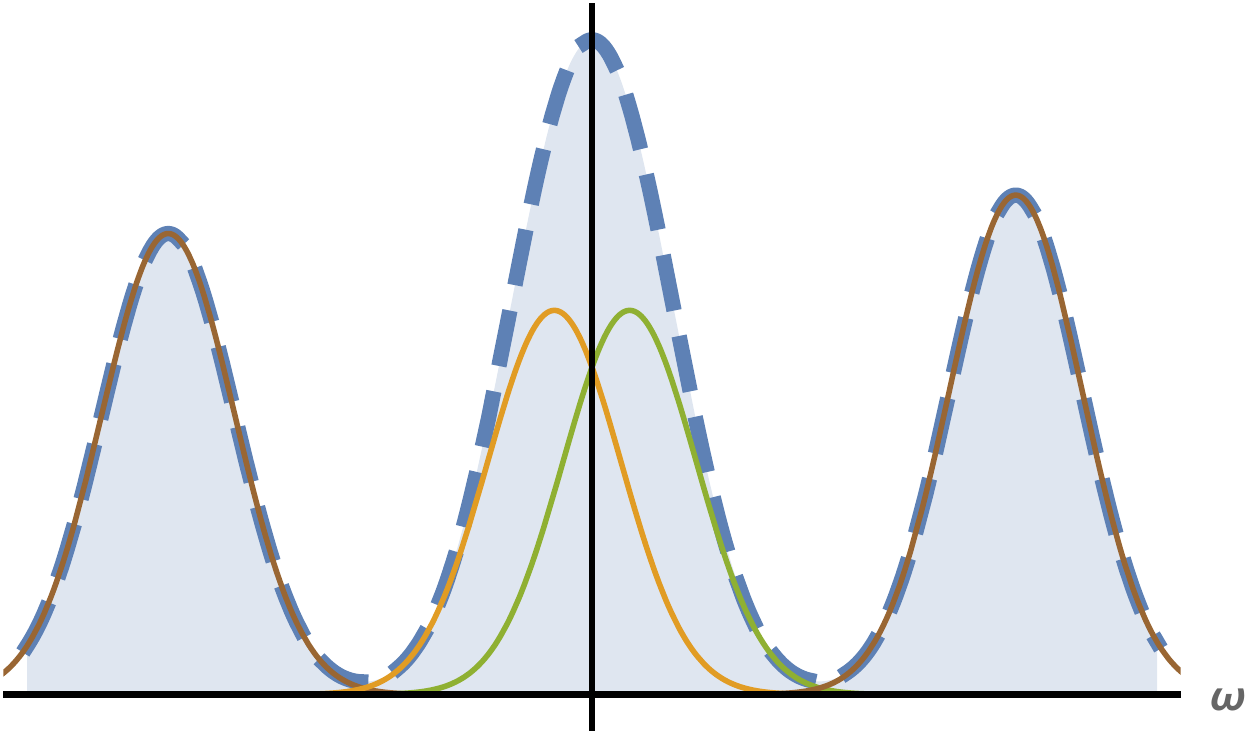}}
\end{center}
\caption{{\bf{A spectrum analysis problem.}} While it is relatively easy to estimate the two side frequencies, the estimation of the two close frequencies is challenging and becomes infinitely difficult when the frequencies merge.}
\label{spect}
\end{figure}

Specifically, this method can be highly useful for analyzing complex spectrums with quantum sensors (such as quantum NMR problems \cite{staudacher2013nuclear,aslam2017nanoscale,pham2016nmr,kong2015towards}).
An example for such a spectrum is illustrated in fig. \ref{spect} . While the two extreme frequencies can easily be estimated, the two central frequencies must be analyzed with a more sophisticated method, which eventually yields higher uncertainty.
 Here, we show that by using a quantum control, the spectrum can be shifted such that the projection noise vanishes. The vanishing projection noise 
implies a finite uncertainty irrespective of the frequency separation. In other words, the uncertainty does not diverge when the two frequencies merge.
 Furthermore, this method is extremely simple, unlike numerically demanding classical superresolution methods.

\section*{Results}
\subsection{Conditions for superresolution}
We first briefly review the pillars of quantum parameter estimation problems. A typical problem involves a quantum state $\rho\left(\theta \right),$ such that $\theta$ is to be estimated.
The uncertainty in estimating $\theta$ is tightly lower bounded by $\frac{1}{\sqrt{I_{\theta}}}$ , where $I_{\theta}$ is the Fisher information (FI) about $\theta$ \cite{cramer2016mathematical}. For a given choice of measurement of $\rho\left( \theta  \right),$ $I_{\theta}$ is determined according to the probabilities $\left(p_{j}\right)$ in the following way: $I_{\theta}=\sum_{j}\frac{\left(\frac{dp_{j}}{d\theta}\right)^{2}}{p_{j}}.$
The FI can be optimized over all possible measurements, leading to the quantum Fisher information (QFI) \cite{wootters1981statistical,braunstein1994statistical}. Given a spectral decomposition $\rho=\sum_{j}p_{j}|\psi_{j}\rangle\langle\psi_{j}|$ the QFI about $\theta$ reads: $\mathcal{F}=\underset{p_{i}+p_{j}\neq0}{\sum}\frac{2}{p_{j}+p_{i}} \bigg \vert  \left(\frac{d \rho}{d \theta}\right)_{i,j} \bigg \vert ^{2}.$

For a multivariable estimation of $\left\{ \theta_{k}\right\} _{k},$ the error is quantified by the covariance matrix of the estimators.
This covariance matrix is lower bounded by $\mathcal{F}^{-1},$ the inverse of the QFI matrix,
 where the QFI matrix is defined as $\mathcal{F}_{k,l}=2\underset{i,j}{\sum}\frac{\left(\frac{\partial\rho}{\partial\theta_{k}}\right)_{i,j}\left(\frac{\partial\rho}{\partial\theta_{l}}\right)_{j,i}}{\left(p_{i}+p_{j}\right)}.$

We are now poised to formulate spectral resolution problems, which are the focus of this paper.
In these problems we are given a signal (Hamiltonian) that oscillates with time.
It consists of at most two frequencies, yet the exact number of frequencies (and their values) is unknown and need to be determined.
To this end, a quantum probe interacts with the signal so that information about it becomes encoded on the probe and can be extracted by measurements.  
Once this information is extracted this problem boils down to a parameter estimation problem: the common strategy in these problems \cite{hannan1989resolution,barabell1998performance,roy1989esprit,rotem2017limits} is to assume that there are two frequencies and estimate them. 
If the estimation shows a significant overlap between the frequencies (significant with respect to the estimation error), it is concluded that the frequencies are not resolvable.
However if the overlap is negligible, one can deduce that the signal consists of two frequencies (since the error probability is negligible). This implies that the figure of merit is $\Delta \omega_{1},$ $\Delta \omega_{2}.$ 
The challenging regime is when $\omega_{1}\rightarrow\omega_{2}.$ Resolution becomes an issue when $\Delta\omega_{1},\:\Delta\omega_{2}\rightarrow\infty$ as $\omega_{1}\rightarrow\omega_{2}.$
A different, and somewhat more convenient, formulation uses $\omega_{ \mathrm{r} }=\frac{\omega_{1}-\omega_{2}}{2},$ $\omega_{ \mathrm{s} }=\frac{\omega_{1}+\omega_{2}}{2},$
so that the resolution condition is $\Delta\omega_{ \mathrm{r} }\ll\omega_{ \mathrm{r}  }$ and the figure of merit is thus $\Delta\omega_{ \mathrm{r} }.$ 
The key issue is thus the behavior of $\Delta\omega_{ \mathrm{r} }$ as $\omega_{ \mathrm{r} }\rightarrow0,$ if $\Delta\omega_{ \mathrm{r}  }\rightarrow\infty$ then a fundamental resolution limit exists which is the case in relevant classical examples \cite{hannan1989resolution,barabell1998performance,roy1989esprit}.

This limitation appears in various resolution problems (not only spectral resolution) and stems from a property of vanishing distinguishability.
Let us define what vanishing distinguishability means.
Given the quantum state of a probe (density matrix $\rho$), that depends on a set of parameters $\left\{ \theta_{i}\right\} _{i},$ the state suffers from a vanishing distinguishability if the set $\left(\frac{\partial\rho}{\partial\theta_{i}}\right)_{i}$ is linear dependent.
An equivalent way to define it: there exists a parameter $g,$ that is a linear combination of $\left\{ \theta_{i}\right\} _{i},$ such that $\frac{\partial\rho}{\partial g}=0.$
Indeed, in many resolution problems as the separation parameter $\omega_{ \mathrm{r} }$ (the difference between the frequencies or , in imaging, the sources) goes to $0,$ there exists a parameter $g$ such that $\frac{\partial\rho}{\partial g}=0.$ 
In this paper we focus on the simplest (yet very common) case that only $\frac{\partial\rho}{\partial\omega_{ \mathrm{r} }}=0$ as $\omega_{ \mathrm{r}}=0.$ In this case $\Delta\omega_{ \mathrm{r} }\rightarrow\infty$ if and only if the FI about $\omega_{\mathrm{r}}$ (denoted as $I_{ \mathrm{r}  }$) vanishes, which implies that $I_{ \mathrm{r} }$ is our figure of merit. 

As an example, consider a signal that acts on a qubit and is given by the following Hamiltonian: 
\begin{equation}
H=\left[ A_{1}\cos\left(\omega_{1}t\right)+B_{1}\sin\left(\omega_{1}t\right)+A_{2}\cos\left(\omega_{2}t\right)+B_{2}\sin\left(\omega_{2}t\right)  \right] \sigma_{z} .
\label{Hamiltonian}
\end{equation}
It is simple to see that this limitation appears whenever the Hamiltonian posses a symmetry for exchange of $\omega_{1}\leftrightarrow\omega_{2}$ (i.e. identical amplitudes).
This symmetry implies a symmetry of $\omega_{ \mathrm{r} }\leftrightarrow-\omega_{  \mathrm{r} },$ 
from which it follows that the state obtained after evolution time $t$ has the same symmetry,$|\psi_{t}\left(\omega_{  \mathrm{r} }\right)\rangle=|\psi_{t}\left(-\omega_{ \mathrm{r}  }\right)\rangle$, and thus $\frac{\partial|\psi_{t}\rangle}{\partial\omega_{  \mathrm{r}  }}=0$ for $\omega_{ \mathrm{r}  }=0.$  
Given the expression of the quantum Fisher information \cite{braunstein1994statistical}, we obtain:
\begin{equation}
I_{  \rm{r}  }\leq4\left[ \left \langle\frac{\partial\psi}{\partial\omega_{ \mathrm{r}  }} \bigg \vert \frac{\partial\psi}{\partial\omega_{  \mathrm{r} }} \right \rangle - \left \vert \left \langle \frac{\partial\psi}{\partial\omega_{ \mathrm{r}  }} \bigg \vert \psi \right \rangle \right \vert ^{2}\right]\rightarrow0,
\label{no_go_1}
\end{equation}
hence resolution is limited. 
Note that applying further control on the probe cannot eliminate this symmetry, and thus cannot remove this resolution limit.

It can be shown that this limitation appears for any quadratures:
there exists a parameter $g$ such that $\frac{\partial H}{\partial g}=0$ for every $t,$ which implies $\frac{\partial|\psi_{t}\rangle}{\partial g}=0$ for every measurement (more details in supplementary note 4).       

So vanishing distinguishability is quite a common property and appears in different resolution problems, but does it always impose a limitation? 

Eq. \ref{no_go_1} shows that whenever the quantum state of the probe ($\rho$) is pure, resolution is limited, however for some mixed states this property does not limit the resolution, 
these are the states that give rise to superresolution: $\frac{d\rho}{d\omega_{\mathrm{r}}}=0$ yet $I_{{\rm r}}\left(\omega_{{\rm r}}\rightarrow0\right)>0.$
A special case of this phenomenon was found and analyzed recently in the context of optical imaging \cite{nair2016far,tsang2016quantum,lupo2016ultimate,chrostowski2017super}(see supplementary note 3).
Can such states be obtained in quantum spectroscopy and various other problems?
In order to understand this, it would be highly desirable to characterize these states and set a sharp condition for superresolution. 

 Let us show that these states can be simply characterized:
  

{\it{Claim:}} Given $\rho\left(\omega_{\mathrm{r}}\right)$ such that $\frac{d\rho}{d\omega_{\mathrm{r}}}=0$ (as $\omega_{\mathrm{r}}\rightarrow0$), then $I_{{\rm r}}\left(\omega_{{\rm r}}\rightarrow0\right)>0$ if and only if at least one of the eigenvalues of $\rho$ goes as $\omega_{\mathrm{r}}^{k},$
where $1<k\leq2$ or equivalently $\frac{d\sqrt{\rho}}{d\omega_{ \mathrm{r} }}\neq0$. The optimal measurement basis converges to an eigenbasis of $\rho$ as $\omega_{ \mathrm{r} }\rightarrow0$.

We briefly illustrate a proof: Given a spectral decomposition $\rho=\underset{j}{\sum}p_{j}|j\rangle\langle j|,$ then:
\begin{equation}
\frac{d\rho}{d\omega_{\rm{r}}}=\underset{j}{\sum}\frac{dp_{j}}{d\omega_{\rm{r}}}|j\rangle\langle j|+i\underset{j,k}{\sum}\left(p_{j}-p_{k}\right)h_{k,j}|k\rangle\langle j|,
\label{derivative_rho}
\end{equation}
where $h$ is a Hermitian operator and $h_{k,j}$ denote its matrix elements in the eigenbasis of $\rho$. Since $\frac{d\rho}{d\omegar} \rightarrow 0$, then for every $j,k$: $\frac{dp_{j}}{d\omegar}\rightarrow0,\;\left(p_{j}-p_{k}\right)h_{k,j}\rightarrow0.$ 
With this notation, the QFI ($\mathcal{F}$) reads (see \cite{braunstein1994statistical}): 
\begin{equation}
\mathcal{F}=\underset{j}{\sum}\frac{\left(\frac{dp_{j}}{d\omegar}\right)^{2}}{p_{j}}+2\underset{j,k}{\sum}\frac{\left(p_{j}-p_{k}\right)^{2}}{p_{j}+p_{k}}|h_{kj}|^{2}.
\end{equation}
The fact that $\left(p_{j}-p_{k}\right)h_{k,j}\rightarrow0$ implies that $\frac{\left(p_{j}-p_{k}\right)^{2}}{p_{j}+p_{k}}|h_{kj}|^{2}\rightarrow 0,$ however $\frac{dp_{j}}{d\omegar}\rightarrow0$ does not imply that $\frac{\left(\frac{dp_{j}}{d\omegar}\right)^{2}}{p_{j}}$ vanishes. It can be seen that given that $\frac{dp_{j}}{d\omegar} \rightarrow 0,$ $\frac{\left(\frac{dp_{j}}{d\omegar}\right)^{2}}{p_{j}}>0$ if and only if there exists $p_{j}\sim\omegar^{k}$ for $1<k\leq2.$      
We then observe that for $\omegar=0,$ $\mathcal{F}\left(\rho\right)=\underset{j}{\sum}\frac{\left(\frac{dp_{j}}{d\omegar}\right)^{2}}{p_{j}},$ which implies that the optimal measurement basis converges to any eigenbasis of $\rho.$

This condition can be shown to be equivalent to $\frac{d\sqrt{\rho}}{d\omegar}\neq0$ (see supplementary note 1). 
It is quite intuitive that one has to demand $\frac{d\sqrt{\rho}}{d\omegar}\neq0,$ since the QFI equals the minimization of all the QFI's of the purifications.  Since purifications go as $\sqrt{\rho},$ $\frac{d\sqrt{\rho}}{d\omegar}=0$ would imply a vanishing derivative of every purification and thus a vanishing QFI.     

This criterion shows that the only way to overcome a vanishing distinguishability is by nullifying the projection noise of one of the outcomes. 

This condition is a special case of a more general (multivariate) criterion.
In the multivariate version  $\left(\frac{\partial\rho}{\partial\theta_{i}}\right)_{i=1}^{n}$ are linearly dependent (with dimension $k<n$) and the relevant question is whether the QFI matrix can be regular. 
 Note that we can choose $\left(\theta_{i}\right)_{i=1}^{n}$ such that $\left(\frac{\partial\rho}{\partial \theta_{i}}\right)_{i=1}^{k}$ are linearly independent and $\frac{\partial\rho}{\partial \theta_{k+1}}=...=\frac{\partial\rho}{\partial \theta_{n}}=0$ ($\theta_{k+1},....,\theta_{n}$ are the problematic parameters).
Then the QFI is regular if and only if the classical FI matrix (i.e. the FI matrix obtained when measuring in the eigenbasis of $\rho$) about the problematic parameters ($\theta_{k+1},....,\theta_{n}$) is regular.
Namely it depends only on the classical FI about these parameters, and thus the optimal measurement basis to estimate these parameters is the eigenbasis of $\rho.$ 
The proof of this condition is quite similar to that of the single variable case, and is given in supplementary note 2.     

Before we move on to applications in quantum sensing, a few remarks are in order:
An accurate formulation of the superresolution condition 
is $\frac{d \rho}{d \omegar} \rightarrow 0$ and 
$\Ir \left( \omegar \rightarrow 0 \right)  >0,$
namely the limit needs to be positive.
That is because we are interested in the behavior of the FI for a very small difference, rather than a vanishing difference.
We mention this point since the FI at $\omega_{r}=0$ can be discontinuous or meaningless (Cramer-Rao bound may be violated), as one of the eigenvalues vanishes \cite{vsafranek2017discontinuities,seveso2019discontinuity}. 
Given a vanishing eigenvalue, the variance of maximum likelihood estimation will vanish (which corresponds to an infinite FI) and thus may not coincide with the limit.       

We also remark that in all cases examined in this paper (as well as in the imaging case) the eigenvalue goes as $\omegar^2.$
Any different power, $1<k<2,$ would in fact lead to a better performance: a divergent FI.  

 

\subsection{Application: spectral resolution without coherence}

\begin{figure}[t]
\begin{center}
\includegraphics[width=9.1 cm]{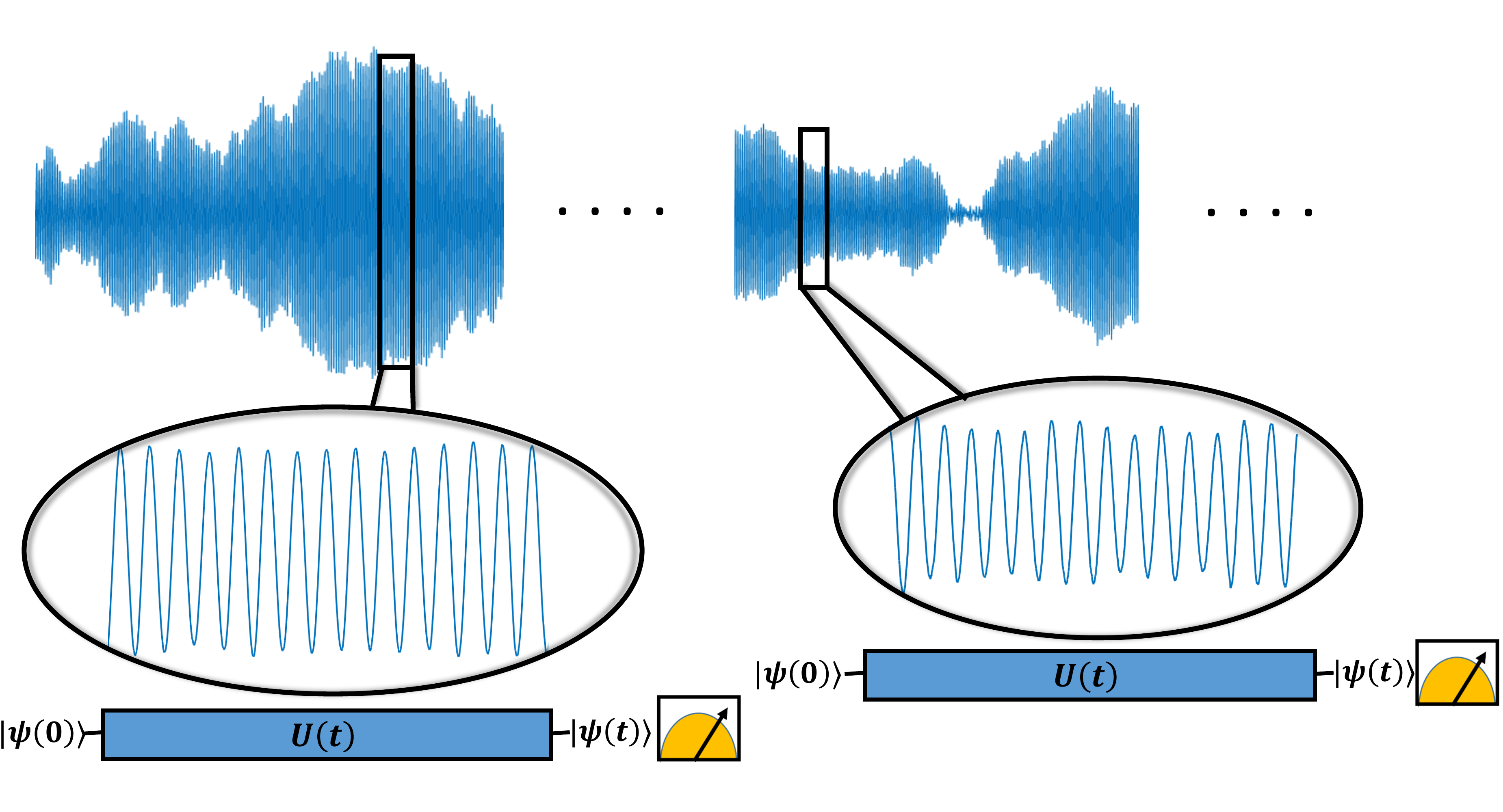}
\end{center}
\caption{ {\bf{The problem of resolution without coherence}}. the quadratures of the signal in different measurements are random. The question we address is whether resolution is limited by the length of individual measurements. }
\label{noisy_signal}
\end{figure}

Consider now again the problem of spectral resolution, with the signal defined in eq. \ref{Hamiltonian}, and such that it suffers from shot-to-shot noise:
in each measurement the frequencies are the same but the quadratures are random, i.e. $A_{i},\:B_{i}$ have a certain distribution. 
Specifically here we assume  $A_{i},\:B_{i}\sim N\left(0,\sigma\right)$, and other noise models are addressed in supplementary note 11.
This scenario is illustrated in fig. \ref{noisy_signal}, and is relevant for different applications, such as communication protocols, spectrum analyzers and nano NMR (\cite{balasubramanian2008nanoscale,gruber1997scanning,maze2008nanoscale,staudacher2013nuclear,mamin2013nanoscale,muller2014nuclear,devience2015nanoscale,aslam2017nanoscale,bucher2018hyperpolarization,lovchinsky2016nuclear,bar2017observing}), in particular when the time required to perform projective measurement is longer than the coherence time of the signal  (this is the case with NV centers, due to the large number of iterations needed, and with trapped ions, where the re-cooling process might be longer than the coherence time of the qubit).


It is quite clear that the fluctuations of the quadratures remove the purity of the probe, which can give rise to superresolution states. Let us examine this.  

Consider a standard Ramsey experiment, in which the probe is initialized in $\sigma_{x}-\sigma_{y}$ plane, then rotated due to the signal and eventually measured in the initialization basis.
Due to the fluctuations of the Hamiltonian, an averaging should be performed.
Therefore the state of the probe is given by a density matrix:
\begin{equation}
 \rho=\int p\left(A_{i}\right)p\left(B_{i}\right)|\psi_{A_{i},B_{i}}\rangle\langle\psi_{A_{i},B_{i}}|\;dA_{i}\:dB_{i},
 \end{equation}
where $|\psi_{A_{i},B_{i}}\rangle$ is the state given a single realization of $A_{i},B_{i}.$   
Note that since the fluctuations are identical:
\begin{equation}
 \rho\left(\omegar\right)=\rho\left(-\omegar\right)  \Rightarrow  \frac{d\rho}{d\omegar}=0  \, \left(\omegar=0\right). 
\end{equation}
Once again, control on the probe does not change this symmetry, hence superresolution can be achieved only if the condition presented above is satisfied: projection noise has to be nullified.
It is therefore desirable to find a measurement scheme that nullifies the projection noise.
It is simple to see that this can be obtained if $ \phi_{A_{i},B_{i}}=0\:\left(\forall A_{i},B_{i}\right),$ where $\phi_{A_{i},B_{i}}$ is the phase accumulated by the sensor (defined as half the rotation angle in the Bloch sphere) per realization, since this implies a vanishing transition probability. 



Our claim is therefore: Given the above noise model, there exist measurement schemes that satisfy the superresolution condition and thus achieve $\Ir>0.$
To see that such methods exist,
observe that the phase accumulated by the sensor given the Hamiltonian in Eq. \ref{Hamiltonian} (when no control is applied) reads:
\begin{equation}
\phi_{A_{i},B_{i}}=\underset{i}{\sum}\frac{A_{i}}{\omega_{i}}\sin\left(\omega_{i}t\right)+\frac{B_{i}}{\omega_{i}}\left(1-\cos\left(\omega_{i}t\right)\right).
 \end{equation} 
Note that given this time evolution the density matrix of the sensor is diagonal in the initialization basis with eigenvalues $p, \, 1-p,$ where $p$ is the average transition probability: $p=\langle\sin\left(\phi_{A_{i},B_{i}}\right)^{2}\rangle_{A_{i},B_{i}}.$
Hence the superresolution condition boils down to $p\sim\omegar^{2}.$
This indeed can be satisfied by simply tuning $t$ such that $\omegas t=2\pi n,$ where $n$ is a non-zero integer.
With this tuning $\phi_{A_{i},B_{i}}=0$ (for $\omegar=0$),
and more specifically : 
 \begin{equation} 
\phi_{A_{i},B_{i}}\approx\frac{\left(A_{1}-A_{2}\right)}{\omegas}\omegar t\rightarrow p\approx\frac{2\sigma^{2}}{\omegas^{2}}\omegar^{2}t^{2},
\label{probability_approximated}
\end{equation}  
Hence the superresolution condition is satisfied and the FI reads:

\begin{equation}
\Ir=\frac{8\sigma^{2}t^{2}}{\omegas^{2}}.
\end{equation}

So nullifying the projection noise indeed cancels the vanishing derivative and a finite $\Ir$ is achieved.

\begin{figure}
\begin{center}
\includegraphics[width=9.4cm]{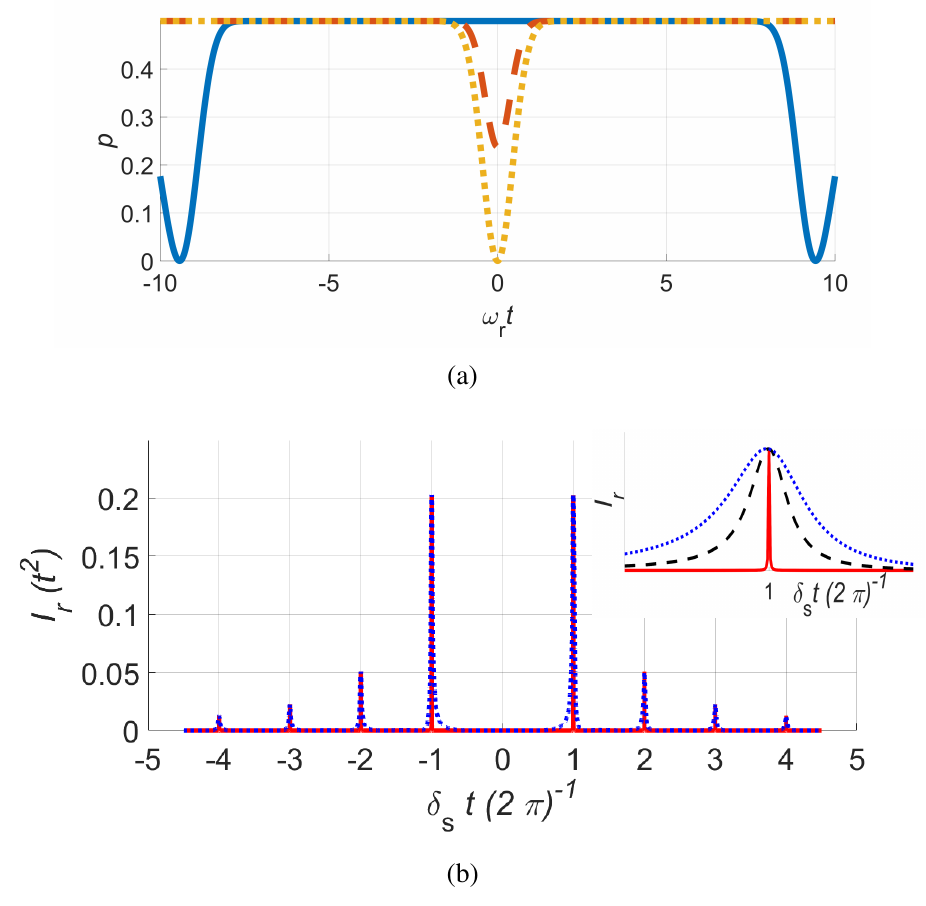}
\end{center}
\caption{ {\bf{Probability and Fisher information analysis}}   (a) Average transition probability ($p$) as a function of $\omegar t$ for different values of $\delta_{ \rm{s} }$: $\delta_{ \rm{s}  }t =\pi$ (blue, solid line), $\delta_{  \rm{s}  }t =1.8 \pi$ (orange, dashed line) and $\delta_{ \rm{s} }t =2 \pi$ (dotted,yellow line). 
For every $\delta_{\rm{s}},$ $\frac{dp}{d\omegar}=0$ for $\omegar=0.$ Hence a finite $\Ir$ can be achieved only if $p=0.$ This requirement is fulfilled when $\delta_{  \rm{s} }t=2\pi n.$
(b) FI about $\omegar$ ($\Ir$) as a function of $\delta_{\rm{s}}t.$ Clear peaks can be observed whenever $\delta_{\rm{s}}t=2\pi n.$ The width of the peaks is illustrated in the inset: for $\omegar=0,$ the width vanishes; however finite $\omegar$ leads to a finite width (given $\omegar t\ll1$ this width goes as $\omegar$, see section \ref{imperfections}). For this illustration: $\omegar t=0.001$ (red, solid line), $\omegar t=0.05$ (black, dashed line),$\omegar t=0.1$ (blue, dotted line).
 }
\label{FI_intro}
\end{figure}

The obtained FI can be still quite poor and far from optimal.
Note that it goes as $1/n^{2},$ where $n$ is the number of periods completed during the measurement. If $n$ is large, then this factor of $\frac{1}{n^{2}}$ can be significant.
 A much better FI can be achieved by applying a suitable control: $\pi-$pulses which effectively change the frequency of oscillations, and reduce $n$ to $1$ \cite{kotler2011single,schmitt2017submillihertz,boss2017quantum,naghiloo2017achievingT4}:

Given an original Hamiltonian of $H=\left[ A\sin\left(\omega t\right)+B\cos\left(\omega t\right)   \right] \sigma_{z},$ applying $\pi-$pulses in a frequency of $\omega+\delta$ (namely a $\pi-$pulse is applied every $\frac{\pi}{\omega+\delta},$ $\delta$ is referred to as detuning) on the probe yields the following effective Hamiltonian (see methods for a derivation): 
\begin{equation}
H_{\text{eff}}=\tan\left(\frac{\pi}{2\left(1+\frac{\delta}{\omega}\right)}\right)\left(\frac{\delta}{\omega}\right)\left[A\sin\left(\delta t\right)+B\cos\left(\delta t\right)\right]\sigma_{z}.
\end{equation}
Hence the $\pi-$pulses effectively change the frequency of the Hamiltonian from $\omega$ to $\delta$ (with a prefactor of $\tan\left(\frac{\pi}{2\left(1+\frac{\delta}{\omega}\right)}\right)\left(\frac{\delta}{\omega}\right)$ added to the amplitude).
Since we aim to reduce the frequency of oscillations, we focus on the limit of $\delta \ll \omega,$  in which $\tan\left(\frac{\pi}{2\left(1+\frac{\delta}{\omega}\right)}\right)\left(\frac{\delta}{\omega}\right)\approx\frac{2}{\pi},$ and thus:
\begin{equation}
H_{\text{eff}}\approx\frac{2}{\pi}\left[A\sin\left(\delta t\right)+B\cos\left(\delta t\right)\right]\sigma_{z}.
\end{equation}       
When dealing with a signal that consists of two frequencies ($\omega_{1},\omega_{2}$), the effective Hamiltonian becomes:          
 \begin{equation}
 H_{\text{eff}} \approx \underset{i}{\sum}\frac{2}{\pi}\left[A_{i}\sin\left(\delta_{i}t\right)+B_{i}\cos\left(\delta_{i}t\right)\right] \sigma_{z}.
 \end{equation}                                    
Hence due to the control the central frequency is shifted to $\delta_{  \mathrm{s}  }=\frac{\delta_{1}+\delta_{2}}{2},$ and the relative frequency simply changes sign: $\delta_{   \mathrm{r}  }=-\omegar.$
The condition of vanishing $p$ becomes: $\delta_{  \mathrm{s}   }t= \pm 2 \pi n,$ such that the optimal strategy is setting $\delta_{ \mathrm{s} }t= \pm 2 \pi .$
Therefore with these (optimal) values of $\delta_{ \mathrm{s}  }$ the FI reads:
\begin{equation}
\Ir \approx \left(\frac{2}{\pi}\right)^{2}\frac{8\sigma^{2}t^{2}}{\delta_{ \mathrm{s}  }^{2}}=\frac{8\sigma^{2}t^{4}}{\pi^{4}}.
\end{equation}
Observe that the scaling of $\Ir$ is optimal (goes as $\sigma^{2} t^{4}$) \cite{pang2017optimal,pang2017optimal, schmitt2017submillihertz, jordan2017classical, yang2017quantum, gefen2017control, naghiloo2017achievingT4}; however it is unknown whether this is the best achievable FI (see extended discussion in supplementary note 6).  
The probabilities and the FI for different detunings are presented in fig. \ref{FI_intro}.  
Note that clear resonance peaks of the FI are observed for $\delta_{ \mathrm{s} }t=\pm 2\pi n,$ any other values of detuning lead to a vanishing FI. 
  
\begin{figure}[t]
\begin{center}
\includegraphics[width=8 cm]{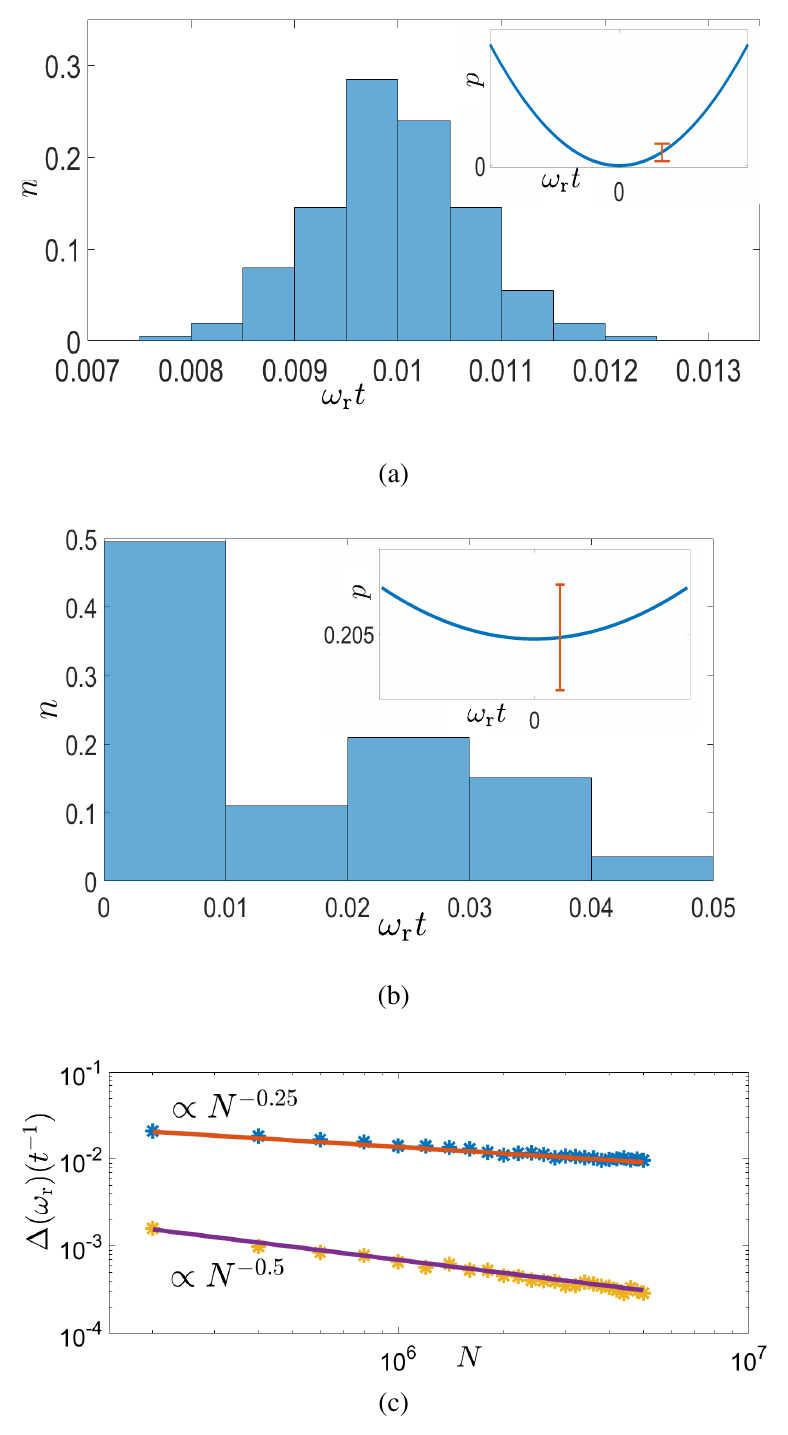}
\end{center}
\caption{   {\bf{Estimation analysis.}}  Maximum likelihood estimation of $\omega_{\rm{r}}$ (beyond the resolution limit; i.e. $\omega_{\rm{r}}t\ll1$) with different control methods.
(a)+(b) Histogram of the estimated $\omega_{\rm{r}}$ for the optimal control method: $\delta_{s}t=2\pi,$ compared to the histogram obtained slightly off the resonance: $\delta_{s}t=1.8 \pi.$
When resonance is achieved, the two frequencies are clearly resolved ($\Delta\omega_{\rm{r}}<\frac{1}{10}\omega_{\rm{r}}$), while off the resonance they are not resolvable ($\Delta\omega_{\rm{r}}>\omega_{\rm{r}}$). Note that off resonance, the standard deviation is too large; hence the probability cannot be distinguished from $p\left(\omega_{\rm{r}}=0\right)$ (see insets).
For both plots $N=10^{6}, \sigma t=5, \omega_{\rm{r}}t=0.01.$
(c) The  RMSE (root mean square error) as a function of $N$ for both control methods. For $\delta_{s}t=2\pi$ the RMSE goes as $\left(NI_{  \rm{r}  }\right)^{-0.5}$ as expected. Off the resonance ($\delta_{s}t=1.8\pi$) the FI vanishes and the RMSE goes as $N^{-0.25}$ (the estimation is biased).         }
\label{numerical_results}
\end{figure}

We tested this method numerically by generating data of two frequency signal (with the corresponding noise model) and performing a Maximum-Likelihood estimation (MLE) to find $\omega_{\rm{r}}.$
Some of the results are shown in fig. \ref{numerical_results}. It can be seen that by choosing a detuning such that $\delta_{  \mathrm{s}  }t=2\pi,$ $\omegar$ can be estimated efficiently and the frequencies are resolved. As shown in fig. \ref{numerical_results}, the standard deviation matches the theoretical expectation: $\Delta\omegar=\frac{1}{\sqrt{I_{  \rm{r}  }N}}.$
By utilizing this control method the number of measurements ($N$) needed to achieve resolution is $N \gg p^{-1}=\frac{\pi^{4}}{2\sigma^{2}\omega_{\rm{r}}^{2}t^{4}}.$ Taking for example values which are well beyond the resolution limit such as $\omega_{\rm{r}}t=0.01,\:\sigma t=1,$ resolution is achieved for $N \gg 5\cdot10^{5}.$  
If the chosen detuning does not satisfy one of theses conditions ($\delta_{s}t=2\pi n$) we expect to observe a divergence in the variance. We used MLE for this case as well. Note that the fact that the FI vanishes does not mean that no information about $\omega_{\rm{r}}$ is obtained, information is in fact obtained from the second derivative.
The estimator becomes biased and the standard deviation reads: $\Delta\omega_{\rm{r}}=\frac{\left(p\left(1-p\right)\right)^{0.25}}{\sqrt{\frac{\partial^{2}p}{\partial\omega_{\rm{r}}^{2}}}N^{0.25}},$ see fig. \ref{numerical_results}. The fact that the standard deviation is proportional to $N^{-0.25}$ (as opposed to the standard scaling of $N^{-0.5}$) is a manifestation of the divergence.    
The resolution condition in this case is thus: $N \gg \frac{p\left(1-p\right)}{\left(\frac{\partial^{2}p}{\partial\omega_{\rm{r}}^{2}}\right)^{2}\omega_{\rm{r}}^{4}}.$ Considering the same example as previously ($\omega_{\rm{r}}t=0.01,\:\sigma t=1$) but with off-resonance detuning $\left(\delta_{s}t=1.8\pi\right)$, 
 the number of measurements required for resolution is $N \gg 10^{8};$ hence a difference of almost three orders of magnitude.        

This method can be understood in the following simple and intuitive way: If there is only a single frequency and $\delta_{s}=\frac{2 \pi}{t}$ then $p_{a}=0,$ hence no transitions should occur, whereas a finite (small) $\omega_{\rm{r}}$ should lead to a small transition probability $\left(p\right)$, such that transitions will be observed after $\frac{1}{p}=\frac{\pi^{4}}{2\sigma^{2}\omega_{\rm{r}}^{2}t^{4}}$ measurements.       

\subsection{Limitations and imperfections}
\label{imperfections}
The method, as analyzed so far, assumes knowledge of all the other parameters ($\sigma$ and $\omega_{\rm{s}}$), coherence of the signal and the probe during the measurement period, and measurements with unit fidelity.  
In this section we analyze each one of these assumptions. The first one to be analyzed is the main caveat of the method: the requirement of coherence during the measurement period.  

\begin{figure}[t]
\begin{center}
\includegraphics[width=8 cm]{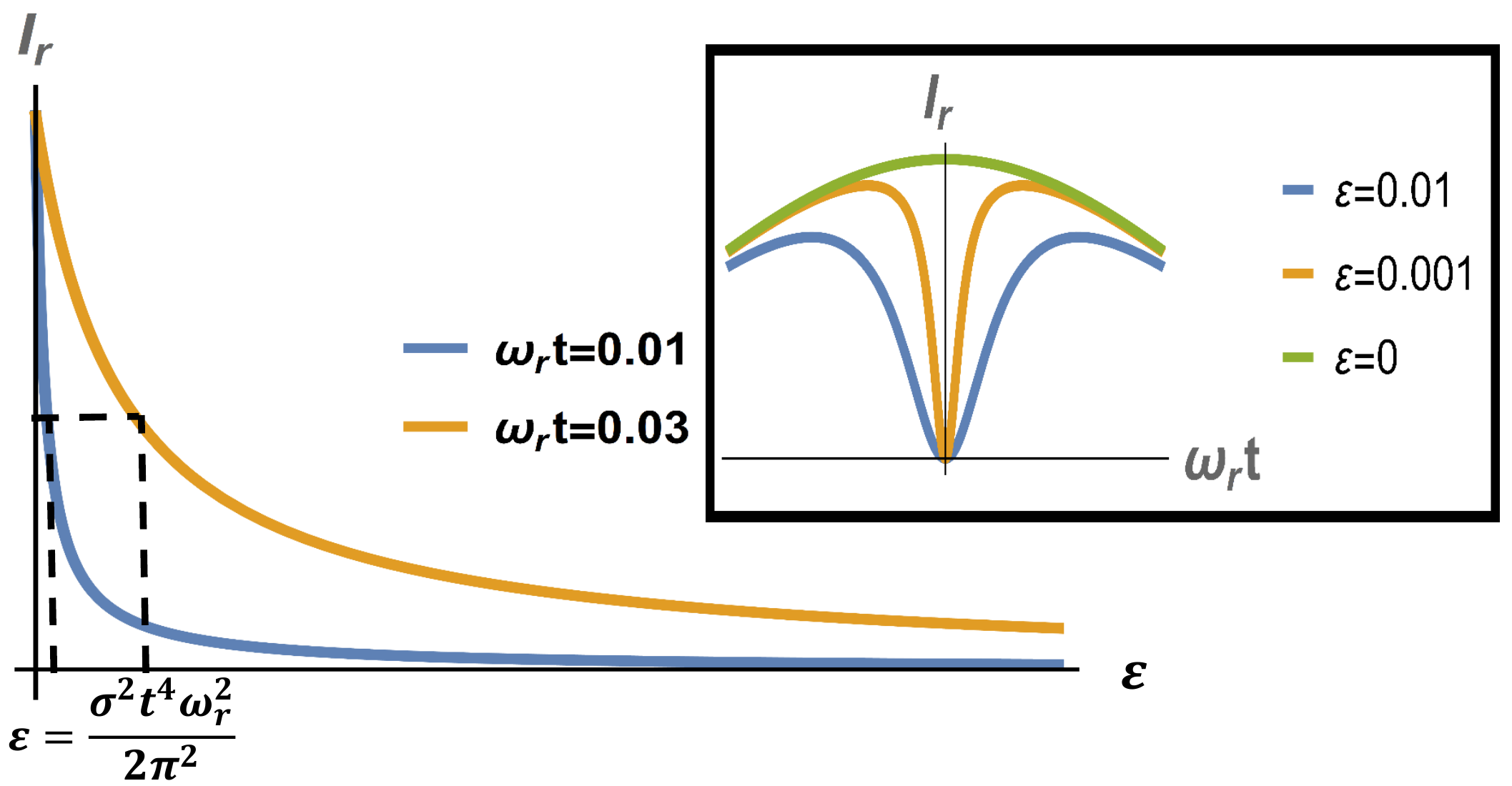}
\end{center}
\caption{ {\bf{The effect of noise.}} $ \Ir$ as a function of $\epsilon$ (a general noise term in the measurement, see eq. \ref{imperfection_probability}) for different values of $\omegar.$ $\Ir$ drops to half the maximal value for $\epsilon=\frac{\left(\sigma t\right)^{2}\left(\omegar t\right)^{2}}{2\pi^{2}},$ 
which means that the maximal $\epsilon$ for which resolution can be achieved goes as $\left(\sigma t\right)^{2}\left(\omegar t\right)^{2}.$
In the inset: $\Ir$ as a function of $\omegar$ for different $\epsilon.$    }
\label{imperfection_plot}
\end{figure}

This method relies on the ability to nullify projection noise, in particular on the fact that for $\omega_{\rm{r}}=0$ the state can become pure.
However this is achieved only for a signal which is perfectly coherent during each measurement.
Fluctuations of the signal during the measurement period inflict a limitation, as in this case it is not possible to nullify the projection noise.
Heuristically due to these fluctuations the transition probability includes an additional noise term (denoted as $\epsilon$), such that it reads (for $\omega_{ \rm{s}  }t=2\pi$):
 \begin{equation}
 p=\frac{\sigma^{2}t^{2}}{2\pi^{2}}\omegar^{2}t^{2}+\epsilon.
 \label{imperfection_probability}
 \end{equation}
This new term imposes a limitation: It is now impossible to nullify $p,$ which implies that $\Ir \rightarrow0$ as $\omegar \rightarrow0.$ The FI $\left(\text{for } \omega_{ \rm{s}  }t=2\pi,\omega_{\rm{r}}t\ll1\right)$ now reads:
\begin{equation}
 I_{{\rm {r}}}\approx\frac{\left(\sigma t\right)^{2}\left(\omega_{\rm{r}}t\right)^{2}}{\pi^{4}\left(\epsilon+\frac{\sigma^{2}t^{2}}{2\pi^{2}}\omega_{\rm{r}}^{2}t^{2}\right)},
\label{FI_imperfection}  
\end{equation}  
this behavior is illustrated in fig. \ref{imperfection_plot},
and it can be observed that resolution can be achieved only for $\omega_{\rm{r}} t>\frac{\sqrt{\epsilon}}{\sigma t}.$

More specifically, assuming a realistic noise model: the quadratures undergo OU (Ornstein-Uhlenbeck) noise process (with variance $\sigma_{n}^{2}$ and damping rate $\gamma$), the noise term reads (in leading order of $\left(\gamma t\ll1\right)$, see supplementary note 8 ) $\epsilon=\frac{\sigma_{n}^{2}t^{3}}{\pi^{2}}.$
When comparing $\epsilon$ to the original transition probability: $\sigma^{2}\frac{t^{2}}{2\pi^{2}}\omega_{\rm{r}}^{2}t^{2}=\frac{\sigma_{n}^{2}t^{2}}{4\pi^{2}\gamma}\omega_{\rm{r}}^{2}t^{2},$ we get the Fourier limit: $\frac{\omega_{\rm{r}}}{\gamma}>1.$ 
We remark that whether one can remove this limitation is an open question.  
Therefore this method is relevant mainly for experimental scenarios with noise that is effectively shot to shot: small enough fluctuations during each measurement but no correlations between consecutive measurements.
This is the case in many experimental settings, where the time separation between measurements is longer than the phase acquisition period due to long readout and preparation stages.      

Quite similarly, dephasing of the probe also imposes a limitation. Taking into account a dephasing rate $\kappa,$ the transition probability reads: $p=0.5\left(1-\exp\left(-\frac{\left(\sigma t\right)^{2}\left(\omega_{\rm{r}}t\right)^{2}}{\pi^{2}}-2\kappa t\right)\right).$
Hence resolution can be achieved only if $ \frac { \omega_{\rm{r}}\sigma}{\left(\kappa^{2}\right)}\gg1.$ Note that in order to retrieve the noiseless FI it is not enough to require $\kappa t \ll 1,$ as there is also minimal time $t>\frac{\kappa^{1/3}}{\sigma^{2/3}\omega_{\rm{r}}^{2/3}}$ ($\kappa t$ should be smaller than $\left(\sigma t\right)^{2}\left(\omega_{\rm{r}}t\right)^{2}$). A detailed analysis of this limitation can be found in supplementary note 9.   

We next address the consequences of imperfect measurements. The effect of imperfect measurements is similar to that of incoherence, therefore the measurement infidelity sets a resolution limit. 
We consider a model in which there are two different outcomes and there is a finite probability to get each outcome from both states (as is the case for the NV center \cite{batalov2008temporal}).           
Namely the probability of detecting an outcome that corresponds to the bright state is: $p=\left(1-\epsilon'\right)p_{b}+\epsilon' p_{d},$ where $p_{b}$ ($p_{d}$) denotes the probability of the bright (dark) state and then $\epsilon'$ is the probability of wrong detection. 
Given this error probability we can observe that $\frac{dp}{d\omegar}=0$ (when $\omegar=0$) but it is impossible to nullify $p$.
This implies  $I_{  \rm{r}  }\rightarrow0$ as $\omegar \rightarrow 0.$
Therefore taking  $\epsilon'\ll1$ (and $\omegar t\ll1,\omega_{\rm{s}}t=2\pi$) we get the same expression as in eq. \ref{imperfection_probability} (with a noise term of $\epsilon'$): $p\approx\frac{\sigma^{2}T^{2}}{2\pi^{2}}\omegar^{2}T^{2}+\epsilon'.$
Hence resolution limit is given by: $\omegar T>\frac{\sqrt{\epsilon'}}{\sigma T}$ (see fig. \ref{imperfection_plot}). 



Let us now address the multivariable estimation protocol. In any realistic scenario $\sigma$ and $\omega_{\rm{s}}$ are unknown. Since the estimation protocol of $\omegar $ depends on knowledge of $\omega_{\rm{s}}$ a preliminary estimation of $\omega_{\rm{s}}$ must be performed
(quite analogously to the preliminary estimation of the centroid in quantum resolution methods for optical imaging \cite{parniak2018beating,nair2016interferometric} ).    
This can be done using the traditional method \cite{pham2016nmr}: Applying $\pi$-pulses in different frequencies and fitting the transition probability as a function of the pulses frequency (see supplementary note 7).
This should provide a good estimation of $\sigma,\omega_{ \rm{s}  },$ but not a good enough estimation of $\omega_{\rm{r}}$ (unless by chance we hit close enough to a resonance frequency).
Once a good enough estimation of $\omega_{ \rm{s}  }$ is obtained we can apply the required control ($\delta_{s}t=2 \pi$). 
To understand what is a good enough estimation of $\omega_{ \rm{s}  }$ observe that for small enough $\omega_{\rm{r}}t,\left( \delta_{s} t-2\pi  \right)$: $I_{  \rm{r}  }=\frac{8\sigma^{2}t^{4}}{\pi^{4}}\frac{\omega_{\rm{r}}^{2}}{\omega_{\rm{r}}^{2}+\left(\delta_{s}-2\pi/t   \right)^{2}},$ hence the width of the resonance peak (in $\delta_{s} t$) goes as $\omega_{\rm{r}}t.$ 
Therefore once $\Delta{\omega_{ \rm{s}  }}$ is comparable to $\omega_{\rm{r}}$ this method works despite the small detuning.

Observe that now a multivariate estimation should be performed, which means that at least three different measurements are needed; each measurement in a detuning that is optimal for a different parameter.
Numerical results and further analysis are presented in supplementary note 7.

\subsection{ Additional applications: quantum resolution methods for sampling }
{ \it{Superresolution with quantum Fourier transform (QFT)} }: Consider the signal (Hamiltonian) in eq. \ref{Hamiltonian}, if the coherence time of the signal is relatively long, its spectrum can be found by sampling it
(where sampling means Ramsey measurements of the probe in different times). Several recent experiments implemented this scheme \cite{schmitt2017submillihertz,boss2017quantum,glenn2018high}. 
The straightforward (and natural) way to analyze this data is by fitting the power spectrum of the measurement outcomes, however it was shown in \cite{rotem2017limits} that this method suffers from a resolution limit.
The reason is again that the (average) power spectrum is symmetric with respect to $\omega_{\rm{r}},$ yet the measurement noise does not vanish.
We point out here that this limit can be eliminated if instead of classical Fourier transform, one uses a quantum Fourier transform. 
In more detail: a phase of $\phi_{j} \approx \tau\underset{i}{\sum}\left(A_{i}\cos\left(\omega_{i}t_{j}\right)+B_{i}\sin\left(\omega_{i}t_{j}\right)\right)$ is accumulated by the probe in each measurement (where $\tau$ is the length of each measurement). 
The idea is that instead of measuring the probe after each phase acquisition, we can map the phases to memory qubits to form the state: $|\psi\rangle=\frac{1}{\sqrt{N}}\underset{j=1}{\overset{N}{\sum}}e^{i\phi_{j}}|j\rangle,$ 
 and then measure in the Fourier basis.
With the appropriate choice of total sampling time ($T=m\frac{2\pi}{\omega_{\rm{s}}}$, for an integer $m$) and measurement length ($\tau=\frac{2\pi}{n\omega_{ \rm{s} }},$ for an integer $n$), the only states in the Fourier basis that can be measured are harmonics of $\omega_{\rm{s}}$ (for $\omegar= 0$),
hence vanishing projection noise of all the others outcomes. The probability to measure the other frequencies, for $\omegar T\ll1,$ is $\approx \frac{1}{6}\omegar^{2}T^{2}\left(\sigma\tau\right)^{2}$ (see supplementary note 10), therefore $\Ir=\frac{2}{3}\left(\sigma\tau\right)^{2}T^{2}$ for $\omegar=0.$      
We thus get a non vanishing FI, and it can be shown that for any model in which the phases are uniformly distributed, the optimal measurement basis is indeed the Fourier basis.

A different method for the same problem is {\it{superresolution with correlation spectroscopy}}: The Hamiltonian is the same, but now perform two measurements and correlate between them using a single memory qubit,
namely the state of the memory qubit after the two phase accumulation periods is $\frac{1}{\sqrt{2}}\left(|0\rangle+e^{i\left(\phi_{1}-\phi_{2}\right)}|1\rangle\right),$ 
where $\phi_{j}$ is the phase accumulated in the $j$th period.
It is simple to see that by choosing the period between measurements to be $T=\frac{2\pi}{\omega_{ \rm{s}  }}n,$ and measuring in the initialization basis the transition probability is $p\approx\langle\left(B_{1}-B_{2}\right)^{2}\rangle\tau^{2}\omega_{\rm{r}}^{2}T^{2}.$ Therefore a non-vanishing FI is achieved:
 $I_{  \rm{r}  }=4\langle\left(B_{1}-B_{2}\right)^{2}\rangle\tau^{2}T^{2}=8\left(\sigma\tau\right)^{2}T^{2}.$

\section*{Discussion}
We presented methods that are capable of resolving frequencies beyond the resolution limits ($\omega_{\rm{r}}t \ll 1$) in quantum spectroscopy.
Those methods are special cases of a general superresolution criterion: one can overcome the vanishing derivative by making the projection noise vanish at the same rate.
The main method that was analyzed (resolution without coherence) is applicable with state of the art experimental capabilities and does not require involved numerical analysis. 

It would be interesting to inquire whether similar ideas are useful to other resolution problems, such as resolving the locations and the frequencies of single neighboring spins.

 The methods presented above are not perfect, they are limited by the noise of the signal and the dephasing of the probe, whether one can overcome these limitations is an open question.

\section*{Methods}

\subsection*{Derivation of density matrix and probabilities}
\label {averaged_probability}
Given a noise model on the amplitudes, the quantum state of the probe is described by the following density matrix:
\begin{equation}
\rho=\int|\psi\rangle\langle\psi|\:p\left(\vect{A},\vect{B}\right)\:d\vect{A}  \,d\vect{B}.
\end{equation}  
Since the time evolution (with and without control) is described by the operator: $U=\cos\left(\phi\right) \mathbb{1} -i\sin\left(\phi\right)\sigma_{z},$
$\rho$ reads: 
\begin{eqnarray}
\begin{split}
&\rho=\int\left[\cos\left(\phi\right)^{2}\rho_{0}+\sin\left(\phi\right)^{2}\sigma_{z}\rho_{0}\sigma_{z}-\frac{i}{2} \sin\left(2 \phi\right) \left[ \sigma_{z},\rho_{0}  \right] \right] \cdot\\ 
&p\left(  \vect{A},  \vect{B}   \right)\:d \vect{A} \,  d\vect{B},
\end{split}
\end{eqnarray}
where $\rho_{0}$ is the initial state. With the  relevant noise model ($A_{i},B_{i}\sim N\left(0,\sigma\right)$) it can be seen that the terms going as $\sin\left(2 \phi\right)$
vanish, leading to: 
\begin{equation}
\rho=\left(1-p\right)\rho_{0}+p\sigma_{z}\rho_{0}\sigma_{z},
\end{equation}
where $p$ is the (averaged) transition probability: $\int\sin\left(\phi\right)^{2}p\left(\vect{A},\vect{B}\right)\:d\vect{A}  \,   d\vect{B}.$
Taking $\phi=\sum_{i}A_{i}\frac{\sin\left(\delta_{i}t\right)}{\delta_{i}}+B_{i}\frac{1-\cos\left(\delta_{i}t\right)}{\delta_{i}},$
a simple calculation yields: 
\begin{equation}
p=0.5\left(1-\exp\left(- 8 \sum_{i}    \frac{\sigma^2}{\delta_i^2}  \sin^2\left(\frac{\delta_i t}{2}  \right)    \right)   \right).
\end{equation}
Note that this expression coincides with eq. \ref{probability_approximated} for $\delta_{ \rm{s} }t=2\pi,$ $\omegar t \ll 1.$
The optimal initial state would be $\rho_{0}=|\uparrow_{x}\rangle\langle\uparrow_{x}|$ (or any other pure state in the $X-Y$ plane), leading to $\rho=\left(1-p\right)|\uparrow_{x}\rangle\langle\uparrow_{x}|+p|\downarrow_{x}\rangle\langle\downarrow_{x}|.$
The QFI (about $\omegar$) of $\rho$ is thus: $\frac{\left(\frac{dp}{d\omegar}\right)^{2}}{p\left(1-p\right)},$
which is the expression of $\Ir$ mentioned in the main text.

\subsection*{Effective Hamiltonian derivation}
\label{effective_hamiltonian_derivation}
In this section we derive the effective Hamiltonian that appears in the main text. 
Given a Hamiltonian: $H=\left[ A \sin\left(\omega t\right) + B\cos\left(\omega t\right)  \right] \sigma_{z},$ and $\pi$-pulses that are applied every $\tau,$ the Hamiltonian in the interaction picture of these pulses is:
\begin{equation}
H=\left[A\sin\left(\omega t\right)+B\cos\left(\omega t\right)\right]h\left(t\right)\sigma_{z},
\end{equation}
where $h \left( t \right)$ is the square wave function.
Note that the phase accumulated by the sensor (denoted as $\phi,$ and defined as half the rotation angle in Bloch sphere) in $t=n \tau$ is:                 
\begin{eqnarray}
\begin{split}
&\phi=A\:\text{Im}\left(\Phi\right)+B\:\text{Re}\left(\Phi\right)\;\text{where:}\\ 
&\Phi=\underset{n=0}{\overset{N-1}{\sum}}\int_{n\tau}^{\left(n+1\right)\tau}e^{i\omega t}\left(-1\right)^{n}\:dt
 \end{split} 
 \end{eqnarray}
where Re (Im) denotes the real (imaginary) part. Therefore in order to find $\phi$ we need to calculate $\Phi$:  
\begin{equation}
\Phi=\underset{n=0}{\overset{N-1}{\sum}}e^{i\omega n\tau}\left(-1\right)^{n}\frac{e^{i\omega\tau}-1}{i\omega}=\underset{n=0}{\overset{N-1}{\sum}}e^{in\left(\omega\tau+\pi\right)}\frac{e^{i\omega\tau}-1}{i\omega}.
\end{equation}
The calculation then proceeds as follows:
\begin{eqnarray}
\begin{split}
&\Phi=\frac{1-e^{iN\left(\omega\tau+\pi\right)}}{1+e^{i\omega\tau}}\frac{e^{i\omega\tau}-1}{i\omega}=\\
&-2 ie^{i\frac{N}{2}\left(\omega\tau+\pi\right)}\sin\left(N\frac{\omega\tau}{2}+N\frac{\pi}{2}\right)\sin\left(\omega\frac{\tau}{2}\right)\frac{1}{\cos\left(\frac{\omega\tau}{2}\right)\omega}.
\end{split}
\end{eqnarray}

Hence: 
\begin{eqnarray}
\begin{split}		
  &\text{Re} \left( \Phi \right)=\left(1-\cos\left(\omega t+N\pi\right)\right)\frac{\sin\left(\omega\frac{\tau}{2}\right)}{\cos\left(\omega\frac{\tau}{2}\right)\omega},\\
  &\text{Im} \left( \Phi  \right)=-\sin\left(\omega t+N\pi\right)\frac{\sin\left(\omega\frac{\tau}{2}\right)}{\omega\cos\left(\frac{\omega\tau}{2}\right)}.
  \label{im_re_expressions}
  \end{split} 
 \end{eqnarray} 
  Note that: $ \omega t=\omega N\frac{\pi}{\omega+\delta}=N\pi-\delta t,$ therefore eq. \ref{im_re_expressions} is simplified to:
 \begin{eqnarray}
\begin{split}		
  &\text{Re}\left(\Phi\right)=\left(1-\cos\left(\delta t\right)\right)\frac{\tan\left(\omega\frac{\tau}{2}\right)}{\omega},\\
  &\text{Im}\left(\Phi\right)=\sin\left(\delta t\right)\frac{\tan\left(\omega\frac{\tau}{2}\right)}{\omega}.
  \label{im_re_expressions_2}
  \end{split} 
 \end{eqnarray}  

 The accumulated phase,$\phi$, thus reads:
 \begin{equation}
 \phi=A\sin\left(\delta t\right)\frac{\tan\left(\omega\frac{\tau}{2}\right)}{\omega}+B\left(1-\cos\left(\delta t\right)\right)\frac{\tan\left(\omega\frac{\tau}{2}\right)}{\omega}.
  \label{full_expression_phase}
  \end{equation} 

Observe that this exact phase is obtained by the following effective Hamiltonian (note that no approximation is used here):
\begin{equation}
H_{\text{eff}}=\tan\left(\omega\frac{\tau}{2}\right)\left(\frac{\delta}{\omega}\right) \left[ A\cos\left(\delta t\right)+B\sin\left(\delta t\right)   \right] 
\sigma_{z},
\end{equation}
hence we can use this effective Hamiltonian to describe the dynamics. This effective Hamiltonian is somewhat similar to the original Hamiltonian in that the frequency is shifted from $\omega$ to $\delta,$ and the amplitude acquires a prefactor of $\tan\left(\omega\frac{\tau}{2}\right)\left(\frac{\delta}{\omega}\right).$ 

 Note that for  $\delta\ll\omega$:
\begin{equation} 
\tan\left(\omega\frac{\tau}{2}\right)\left(\frac{\delta}{\omega}\right)=\tan\left(\frac{\pi}{2\left(1+\frac{\delta}{\omega}\right)}\right)\left(\frac{\delta}{\omega}\right)\approx\frac{2}{\pi},
\end{equation}    
which implies:
\begin{equation}  
H_{\text{eff}}\approx \left[  A\left(\frac{2}{\pi}\right)\cos\left(\delta t\right)+B\left(\frac{2}{\pi}\right)\sin\left(\delta t\right) \right]  \sigma_{z}     \; \; \left(\delta\ll\omega\right).
\end{equation}
It should be noted that this is the relevant regime for experimental realizations \cite{glenn2018high,laraoui2013high,staudacher2013nuclear,lovchinsky2016nuclear}.
Similarly for the opposite limit $\left(\delta\gg\omega\right),$ we obtain that:
\begin{equation}
H_{\text{eff}}\approx \left[ A\left(\frac{\pi}{2}\right)\cos\left(\delta t\right)+B\left(\frac{\pi}{2}\right)\sin\left(\delta t\right)  \right]  \sigma_{z}  \; \; \left(\omega\ll\delta\right).
\end{equation}

 This can of course be trivially extended for a signal that consists of two frequencies.

\vspace{2mm}
{\bf{Data availability:}} The code and data used in this work are available on request to the corresponding author.

T

\vspace{2mm}
{\bf{Correspondence:}} Correspondence and request for materials should be addressed to: tuvia.gefen@mail.huji.ac.il

\vspace{2mm}
{\bf{ Acknowledgements:}}  This project has received funding from the European Union Horizon 2020 research and innovation
programme ERC grant QRES under grant agreement No 770929 and the collaborative European project  ASTERIQS. 
T.G. is supported by the Adams Fellowship Program of the Israel Academy of Sciences and Humanities 


%

\begin{widetext}

\appendix

\begin{center}
{\bf Supplementary Material}
\end{center}

\section{Conditions for quantum resolution}
\label{sec:resolution_criterion}
The following claim is stated in the main text: Given $\rho\left(\omega_{ \rm{r}  }\right)$ such that $\frac{d\rho}{d\omega_{ \rm{r}  }}=0$ (as $\omega_{ \rm{r}  }\rightarrow0$), then $I_{{\rm r}}\left(\omega_{r}\rightarrow0\right)>0$ if and only if one of the eigenvalues of $\rho$ goes as $\omega_{ \rm{r}  }^{k}$ for $1<k \leq 2$
, or equivalently if and only if $\frac{d\sqrt{\rho}}{d\omega_{ \rm{r}  }}\neq0$. The optimal measurement basis converges to an eigenbasis of $\rho$ as $\omega_{ \rm{r}  }\rightarrow0$.
We showed in the main text that the first condition (at least one of the eigenvalues $\sim \omega_{ \rm{r}  }^{1<k \leq 2}$) is sufficient and necessary.

 First, let us clarify one point: 
$\frac{dp_{j}}{d\omega_{ \rm{r}  }}$ is defined as the derivative of the j-th eigenvalue (at $\omega_{ \rm{r}  }=0$), note that it equals the derivative of the probability of the j-th eigenstate (at $\omega_{ \rm{r}  }=0$), to see this:
\begin{equation}   
\frac{d}{d\omega_{ \rm{r}  }}\langle\psi_{j}|\rho|\psi_{j}\rangle|_{\omega_{ \rm{r}  }=0}=\langle\psi_{j}\left(0\right)|\frac{d\rho}{d\omega_{ \rm{r}  }}|\psi_{j}\left(0\right)\rangle+\langle\frac{d\psi_{j}}{d\omega_{ \rm{r}  }}|\rho\left(0\right)|\psi_{j}\left(0\right)\rangle+\langle\psi_{j}\left(0\right)|\rho\left(0\right)|\frac{d\psi_{j}}{d\omega_{ \rm{r}  }}\rangle=\langle\psi_{j}\left(0\right)|\frac{d\rho}{d\omega_{ \rm{r}  }}|\psi_{j}\left(0\right)\rangle,
\end{equation}
where $|\psi_{j}\left(\omega_{ \rm{r}  }\right)\rangle$ is the $j$-th eigenstate of $\rho\left(\omega_{ \rm{r}  }\right).$ The second equality is due to: $\langle\frac{d\psi_{j}}{d\omega_{ \rm{r}  }}|\psi_{j}\rangle+\langle\psi_{j}|\frac{d\psi_{j}}{d\omega_{ \rm{r}  }}\rangle=0.$  
Therefore the eigenbasis of $\rho$ attains the QFI \Big($\underset{j}{\sum}\frac{\left(\frac{dp_{j}}{d\omega_{ \rm{r}  }}\right)^{2}}{p_{j}}$ \Big). 

An alternative way to see that the eigenbasis is an optimal measurement basis proceeds as follows:
\begin{equation}
\frac{d\rho}{d\omega_{ \rm{r}  }}=0\Rightarrow L\rho+\rho L=0,
\end{equation}
where $L$ is the symmetric logarithmic derivative operator (its eigenbasis is the optimal measurement basis \cite{braunstein1994statistical}) .
Note that the fact that $\rho,L$ anticommute impliess that $L\rho=\rho L=0,$ because given that $|l\rangle$ is an eigenstate of $L$ with an eigenvalue $l \neq 0,$ then: 
\begin{equation}
\langle l |\left(L\rho+\rho L\right)| l \rangle=\left(l+l^{*}\right)\langle l |\rho| l \rangle=0 \Rightarrow \langle l |\rho| l \rangle=0\Rightarrow\rho| l \rangle=0.
 \end{equation} 
 So taking an eigenbasis of $L$: $\left\{ |l_{1}\rangle,...,|l_{n}\rangle\right\} $ with eigenvalues $l_{1},...,l_{n}.$ It can be seen that $\forall i\;\rho L|l_{i}\rangle=0$ and thus $L\rho=\rho L=0.$  
 Therefore they have a common eigenbasis and it attains the QFI.

We next show that the second condition is sufficient and necessary. First, one can see directly that the two conditions are equivalent. Using the same notations as in the main text, we can simply find $\frac{d\sqrt{\rho}}{d\omega_{ \rm{r}  }}:$      
\begin{equation}
\frac{d\sqrt{\rho}}{d\omega_{ \rm{r}  }}=\underset{j}{\sum}\frac{\frac{dp_{j}}{d\omega_{ \rm{r}  }}}{2\sqrt{p_{j}}}|j\rangle\langle j|+i\underset{j,k}{\sum}\left(\sqrt{p_{j}}-\sqrt{p_{k}}\right)h_{k,j}|k\rangle\langle j|.
\label{derivative_rho}
\end{equation}
Since $\frac{d\rho}{d\omega_{ \rm{r}  }}=0$ then $\left(p_{j}-p_{k}\right)h_{k,j}=0 \; (\forall k,j)$ therefore $\left(\sqrt{p_{j}}-\sqrt{p_{k}}\right)h_{k,j}=0.$
While $\frac{\frac{dp_{j}}{d\omega_{ \rm{r}  }}}{2\sqrt{p_{j}}}\neq0$ if and only if $p_{j}\sim\omega_{ \rm{r}  }^{k} \; \left( 1<k\leq2  \right).$ Therefore these conditions are equivalent. 

In fact the more general statement is: for any $\rho\left(\theta\right)$ the QFI (about $\theta$) vanishes if and only if $\frac{d\sqrt{\rho}}{d\theta}=0.$

This fact is a simple coclusion of the following claim (which we prove):\\
{\bf{Claim:}} The QFI ($\mathcal{F}$) about $\theta$ satisfies:
\begin{equation}
 2 \, \text{trace}\left[\left(\frac{d\sqrt{\rho}}{d\theta}\right)^{2}\right]\leq\mathcal{F}\leq4 \, \text{trace}\left[\left(\frac{d\sqrt{\rho}}{d\theta}\right)^{2}\right].
\end{equation}

Proof: using supplementary equation \ref{derivative_rho} we get:
\begin{equation}
\text{trace}\left[\left(\frac{d\sqrt{\rho}}{d\theta}\right)^{2}\right]=\underset{j,k}{\sum}\left(\frac{d\sqrt{\rho}}{d\theta}\right)_{j,k}\left(\frac{d\sqrt{\rho}}{d\theta}\right)_{k,j}=\underset{j}{\sum}\frac{\left(\frac{dp_{j}}{d\theta}\right)^{2}}{4p_{j}}+\underset{j,k}{\sum}\left(\sqrt{p_{j}}-\sqrt{p_{k}}\right)^{2}|h_{j,k}|^{2}.
\end{equation}
Recall that $\mathcal{F}$ reads:
\begin{equation}
\mathcal{F}=\underset{j}{\sum}\frac{\left(\frac{dp_{j}}{d\omega_{ \rm{r}  }}\right)^{2}}{p_{j}}+2\underset{j,k}{\sum}\frac{\left(p_{j}-p_{k}\right)^{2}}{p_{j}+p_{k}}|h_{kj}|^{2}
\end{equation} 
Now observe that: 
\begin{equation}
\frac{\left(p_{j}-p_{k}\right)^{2}}{p_{j}+p_{k}}=\frac{\left(\sqrt{p_{j}}-\sqrt{p_{k}}\right)^{2}\left(\sqrt{p_{j}}+\sqrt{p_{k}}\right)^{2}}{p_{j}+p_{k}}=\left(\sqrt{p_{j}}-\sqrt{p_{k}}\right)^{2}\left[1+\frac{2\sqrt{p_{j}}\sqrt{p_{k}}}{p_{j}+p_{k}}\right].
\end{equation}
Therefore:
\begin{equation}
\left(\sqrt{p_{j}}-\sqrt{p_{k}}\right)^{2}\leq\frac{\left(p_{j}-p_{k}\right)^{2}}{p_{j}+p_{k}}\leq2\left(\sqrt{p_{j}}-\sqrt{p_{k}}\right)^{2}.
\end{equation}
So on one hand:
\begin{equation}
4 \, \text{trace}\left[\left(\frac{d\sqrt{\rho}}{d\theta}\right)^{2}\right]=\underset{j}{\sum}\frac{\left(\frac{dp_{j}}{d\theta}\right)^{2}}{p_{j}}+4\underset{j,k}{\sum}\left(\sqrt{p_{j}}-\sqrt{p_{k}}\right)^{2}|h_{j,k}|^{2}\geq\underset{j}{\sum}\frac{\left(\frac{dp_{j}}{d\omega_{ \rm{r}  }}\right)^{2}}{p_{j}}+2\underset{j,k}{\sum}\frac{\left(p_{j}-p_{k}\right)^{2}}{p_{j}+p_{k}}|h_{kj}|^{2}=\mathcal{F},
\end{equation}
and on the other hand:
\begin{equation}
2 \text{trace}\left[\left(\frac{d\sqrt{\rho}}{d\theta}\right)^{2}\right]=\underset{j}{\sum}\frac{\left(\frac{dp_{j}}{d\theta}\right)^{2}}{2p_{j}}+2\underset{j,k}{\sum}\left(\sqrt{p_{j}}-\sqrt{p_{k}}\right)^{2}|h_{j,k}|^{2} \leq \mathcal{F}.
\end{equation}
Combining the last two inequalities we get the desired inequality: $2 \, \text{trace}\left[\left(\frac{d\sqrt{\rho}}{d\theta}\right)^{2}\right]\leq   \mathcal{F}  \leq 4 \, \text{trace}\left[\left(\frac{d\sqrt{\rho}}{d\theta}\right)^{2}\right].$

Note that the lower bound is saturated for pure states (where only the quantum state is changed) and the upper bound for cases in which only the eigenvalues are changed.

{\bf{Conclusion:}} the QFI vanishes if and only if $\frac{d\sqrt{\rho}}{d\theta}$ vanishes.

The proof is immediate: $\left(\frac{d\sqrt{\rho}}{d\theta}\right)^{2}$ is a positive semidefinite, Hermitian operator. 
Therefore the trace vanishes if and only if $\left(\frac{d\sqrt{\rho}}{d\theta}\right)^{2}$ vanishes, that vanishes if and only if $\frac{d\sqrt{\rho}}{d\theta}$ vanishes.

\section{Conditions for superresolution: multivariable case}
\label{sec:multivariable_criterion}
We wish to prove the condition for a non-singular QFI given that $\left(\frac{\partial\rho}{\partial \theta_{i}}\right)_{i=1}^{n}$ are linear dependent (with dimension $k<n$).
Let us set the stage for the statement.
We can choose the parameters $\left(\theta_{i}\right)_{i=1}^{n}$ such that $\left(\frac{\partial\rho}{\partial \theta_{i}}\right)_{i=1}^{k}$ are linear independent, and $\frac{\partial\rho}{\partial \theta_{k+1}}=...=\frac{\partial\rho}{\partial \theta_{n}}=0.$

{\bf{Definition:}} the classical FI matrix of $\rho$ is the FI matrix according to the eigenvalues of $\rho$ (namely the FI matrix achieved when measuring in the eigenbasis of $\rho$).

The claim is that the QFI matrix is non-singular if and only if the classical FI matrix about the subset $\left\{ \theta_{i}\right\} _{i=k+1}^{n}$ is non-singular.    

To prove this claim we use some facts in quantum and classical estimation theory.
Recall that the QFI matrix (denoted as $\mathcal{F}$) reads:
\begin{equation}
\mathcal{F}_{m,l}=2\underset{i,j}{\sum}\frac{\left(\frac{\partial\rho}{\partial \theta_{m}}\right)_{i,j}\left(\frac{\partial\rho}{\partial \theta_{l}}\right)_{j,i}}{\left(p_{i}+p_{j}\right)},
\end{equation}    
where $p_{j}$ is the j-th eigenvalue of $\rho,$ and the matrix elements are in the eigenbasis of $\rho.$
inserting supplementary equation \ref{derivative_rho} , we get that: 
\begin{equation}
\mathcal{F}_{m,l}=\underset{j}{\sum}\frac{\left(\frac{\partial p_{j}}{\partial \theta_{m}}\right)\left(\frac{\partial p_{j}}{\partial \theta_{l}}\right)}{p_{j}}+\underset{i,j}{\sum}\frac{\left(p_{i}-p_{j}\right)^{2}}{p_{i}+p_{j}}\left(h_{i,j}^{m}h_{j,i}^{l}+h_{j,i}^{m}h_{i,j}^{l}\right),
\end{equation}  
where $h^{m}$ is the Hermitian operator that corresponds to $\frac{\partial}{\partial \theta_{m}}.$
Note that just like in the single-variable case, the first term is the information that we gain from the change in the eigenvalues and the second term is the information that we gain from the change in the eigenvectors.
The first term is thus the classical FI matrix (defined earlier), and is denoted from now on as $C.$ 
The second term can be thought of as the quantum part of the QFI, and is denoted from now on as $Q.$
For convenience let us split $C$ and $Q$ into blocks according to $\left\{ \theta_{i}\right\} _{i=1}^{k}$ and $\left\{ \theta_{i}\right\} _{i=k+1}^{n}$  :
\begin{equation}
C=\left(\begin{array}{cc}
C^{11} & C^{12}\\
C^{21} & C^{22}
\end{array}\right),\;Q=\left(\begin{array}{cc}
Q^{11} & Q^{12}\\
Q^{21} & Q^{22}
\end{array}\right) ,  
\end{equation}
where $C^{11}$ is the classical FI about $\left\{ \theta_{i}\right\} _{i=1}^{k},$ $C^{22}$ is the classical FI about $\left\{ \theta_{i}\right\} _{i=k+1}^{n}$ (and analogously for Q).

{\bf{Claim 1:}} Given that $\frac{\partial\rho}{\partial \theta_{k+1}}=...=\frac{\partial\rho}{\partial \theta_{n}}=0,$ then $Q^{22}=0,\,Q^{12}=Q^{21}=0.$

proof: Just like in the single-variable case, $\frac{\partial\rho}{\partial \theta_{m}}=0$ implies ($\forall i,j$) $\left(p_{i}-p_{j}\right)h_{i,j}^{m}=0$ and therefore ($\forall l$) $\frac{\left(p_{i}-p_{j}\right)^{2}}{p_{i}+p_{j}} h_{i,j}^{m}h_{j,i}^{l}=0$ (because $\frac{\left(p_{i}-p_{j}\right)^{2}}{p_{i}+p_{j}}|h_{i,j}^{m}|\leq\left(p_{i}-p_{j}\right)|h_{i,j}^{m}|\rightarrow0$).        
This implies that $Q^{22},Q^{12},Q^{21}$ vanish.

{\bf{Claim 2:}} Given that $C^{22}$ is singular, then $C$ is singular, and vectors that nullify $C^{22}$ nullify also $C$.

proof: Recall that a (classical) FI matrix is defined as:
\begin{eqnarray}
\begin{split}
&I_{m,l}=\underset{i}{\sum}\frac{\left(\frac{\partial p_{i}}{\partial \theta_{m}}\right)\left(\frac{\partial p}{\partial \theta_{l}}\right)}{p_{i}}=4\underset{i}{\sum}\left(\frac{\partial\sqrt{p_{i}}}{\partial \theta_{m}}\right)\left(\frac{\partial\sqrt{p_{i}}}{\partial \theta_{l}}\right)\\
&=\langle\frac{\partial\sqrt{p}}{\partial \theta_{m}},\frac{\partial\sqrt{p}}{\partial \theta_{l}}\rangle.
\end{split} 
 \end{eqnarray} 
 So it is an inner product matrix between the vectors $\left\{ \frac{\partial\sqrt{p}}{\partial \theta_{m}}\right\} _{m=1}^{n}.$
 Hence it is regular if and only if these vectors are linear independent, and the null-space is all the linear combinations of these vectors that vanish.
 Therefore, if $C^{22}$ is singular then  $\left\{ \frac{\partial\sqrt{p}}{\partial \theta_{m}}\right\} _{m=k+1}^{n}$ is linear dependent which implies that $\left\{ \frac{\partial\sqrt{p}}{\partial \theta_{m}}\right\} _{m=1}^{n}$ is linear dependent,
 and the null-space of $C^{22}$ is a subspace of the null-space of $C.$ 

This immediately leads to the desired conclusion:

{\bf{Conclusion:}} Given that $\left(\frac{\partial\rho}{\partial \theta_{i}}\right)_{i=1}^{k}$ are linear independent and $\frac{\partial\rho}{\partial \theta_{k+1}}=...=\frac{\partial\rho}{\partial \theta_{n}}=0$ (the problematic parameters),
then the QFI matrix is regular if and only if $C^{22}$ (the classical FI about the problematic parameters) is regular.

Proof: 
We first show that if $C^{22}$ is singular then the QFI is singular.
$C^{22}$ is singular and is thus nullified by a vector $\vect{\alpha}$. From claim 2, this $\vect{\alpha}$ nullifies also $C,$ and from claim 1 it nullifies also $Q.$
Therefore the QFI matrix is nullified by $\vect{\alpha}$ and is thus singular.    

We now show that if the QFI is singular then $C^{22}$ is singular.
Given that the QFI is nullified by $\vect{\alpha},$ then $2\underset{i,j}{\sum}\frac{|\bm{\alpha}\cdot\left(   \bm {\partial_{\theta}  \rho  }   \right)_{i,j}|^{2}}{\left(p_{i}+p_{j}\right)}=0.$ 
and therefore $\vect {\alpha}\cdot\left(  \bm{    \partial_{\theta} \rho   }    \right)=0.$
This means that $\alpha_{1}=...=\alpha_{k}=0,$ namely this vector is a linear combination of only the problematic parameters and thus $C^{22} \vect {\alpha}=0.$    
Hence $C^{22}$ is singular. $\square$

As it is mentioned in a footnote, one can formulate an equivalent condition.
Recall that in the single variable case the QFI is positive $\Leftrightarrow$ $\frac{d\sqrt{\rho}}{d\theta}\neq0.$ 
An immediate conclusion of this is that in the multivariable case the QFI matrix is non-singular $\Leftrightarrow$ $\left\{ \frac{\partial\sqrt{\rho}}{\partial\theta_{i}}\right\} _{i=1}^{n}$ are linear independent:
the QFI matrix is singular $\Leftrightarrow$ there exist a parameter $y,$ a linear combination of $\left(\theta_{i}\right)_{i},$ such that the (single-variable) QFI about $y$ vanishes 
$\Leftrightarrow$ $\frac{d\sqrt{\rho}}{dy}=0$ $\Leftrightarrow$ $\left\{ \frac{\partial\sqrt{\rho}}{\partial\theta_{i}}\right\} _{i=1}^{n}$ are linear dependent.

\section{Relation to quantum superresolution in imaging}
\label{sec:imaging}
In this part we revisit the recent superresolution scheme proposed in \cite{tsang2016quantum} and show that it is a special case of the criterion presented in the main text. In the imaging problem one has two close incoherent optical sources, located in $x_{1},x_{2},$ and the goal is to determine the number of sources (two or one) and estimate their positions.
The probe in this problem is the radiation emitted from the sources (detected by a measurement device). In the far-field limit, all terms higher than single photon terms can be neglected, so that the state of the radiation reads:
\begin{equation}
\rho=\left(1-\epsilon\right)|\text{vac}\rangle\langle\text{vac}|+\epsilon\rho_{1}+O\left(\epsilon^{2}\right),\;\text{where}\;\rho_{1}=\frac{1}{2}\left(|\psi_{1}\rangle\langle\psi_{1}|+|\psi_{2}\rangle\langle\psi_{2}|\right),  
\end{equation}
and $|\psi_{j}\rangle=\int\psi_{j}\left(x\right)|1,x\rangle\,dx$ ($|1,x\rangle$ is the state of one photon in position $x$).
$|\psi_{j}\left(x\right)\rangle$ is the photonic wave function corresponding to the $j$-th source, we consider the symmetric case in which $\psi_{j}\left(x\right)=\psi\left(x-x_{j}\right)$ where $x_j$ is the position of the j-th source.

We can now define the parameters $\theta_{1}=\frac{1}{2}\left(x_{1}+x_{2}\right)$ (the centroid, equivalent to $\omega_{s}$ in our case), and $\theta_{2}=x_{1}-x_{2}$ (the distance between sources, equivalent to $\omega_{r}$ in our case).
Replacing $x_{1}\longleftrightarrow x_{2}$ leads to $\psi_{1}\left(x\right)\longleftrightarrow\psi_{2}\left(x\right)$ which does not change $\rho.$
Hence in this case, $\rho$ is symmetric with respect to the parameter $\theta_{2}$ namely $\rho\left( \theta_{2} \right)=\rho\left( -\theta_{2}      \right),$ and thus:
\begin{equation}
\frac{\partial\rho}{\partial\theta_{2}}\rightarrow0,\;\theta_{2}\rightarrow0.
\end{equation}  
Therefore $\rho$ suffers from a vanishing distinguishability.
 According to the criterion in the main text, the optimal measurement basis would converge to an eigenbasis of $\rho$ and a finite FI about $\theta_{2}$ can be achieved if and only if one of the eigenvalues $\rightarrow 0$ as $\theta_{2}^{2}.$    
Observe that the eigenstates of $\rho$ (in the subspace of one photon states) are $|\psi_{1}\rangle\pm\frac{\langle\psi_{2}|\psi_{1}\rangle}{|\langle\psi_{2}|\psi_{1}\rangle|}|\psi_{2}\rangle$ with eigenvalues $\frac{\epsilon}{2}\left(1\pm|\langle\psi_{2}|\psi_{1}\rangle|\right)$ respectively.
Therefore the condition is satisfied given that $1-|\langle\psi_{1}|\psi_{2}\rangle| \sim \theta_{2}^2$ as $\theta_{2}\rightarrow0,$ which is the case for a wide variety of $\psi\left(x\right)$ (e.g. Gaussian, sinc functions and many more). 
Hence the superresolution method in this case, as proposed in \cite{tsang2016quantum} and according to the criterion in the main text, is to measure whether the one photon state is in $|\psi\left(x\right)\rangle.$ The probability of not being in this state goes as $\theta_{2}^{2}$ and thus a finite FI is achieved in the limit of $\theta_{2}\rightarrow0.$

Note that given a symmetric pure state (unnormalized): $|\psi_{1}\rangle+|\psi_{2}\rangle$ resolution cannot be achieved (due to purity) just like the resolution limit in spectroscopy given a coherent symmetric signal.
The state $\frac{1}{2}\left(|\psi_{1}\rangle\langle\psi_{1}|+|\psi_{2}\rangle\langle\psi_{2}|\right)$ is the ensemble average of states with random phases: 
$|\psi_{1}\rangle+e^{i\phi}|\psi_{2}\rangle,$ which is similar to averaging over many realizations of random phase signals that we make in spectroscopy.

\section{Resolution limitations in quantum spectroscopy}
\label{sec:spectroscopy}
It is shown in the main text that given the following Hamiltonian (identical quadratures):
\begin{equation}
H=\left[\underset{i}{\sum}A\cos\left(\omega_{i}t\right)+B\sin\left(\omega_{i}t\right)\right]\sigma_{z}=\left[\underset{i}{\sum}\Omega\sin\left(\omega_{i}t+\varphi\right)\right]\sigma_{z} ,
\label{Hamiltonian2}
\end{equation}  
$(\Omega=\sqrt{A^{2}+B^{2}}, \varphi=\arctan\left(\frac{B}{A}\right)),$ then:

\begin{equation}      
\frac{\partial|\psi_{t}\rangle}{\partial\omega_{ \rm{r}  }}=0\Rightarrow I_{ \rm{r}  }=0\:\left(\omega_{ \rm{r}  }=0\right).
\end{equation}
Note that a similar limitation appears also in the case of identical phases but different amplitudes:
\begin{equation}
H=\underset{i}{\sum}\Omega_{i}\sin\left(\omega_{i}t+\varphi\right)\sigma_{z}.
\end{equation}
Of course, $\frac{\partial H}{\partial\omega_{ \rm{r}  }}\neq0,$ however neither $\omega_{ \rm{s}  }$ nor $\omega_{ \rm{r}  }$ can be efficiently estimated: To see this observe that:
\begin{equation}
\forall t\;\Omega_{2}\frac{\partial H}{\partial\omega_{1}}=\Omega_{1}\frac{\partial H}{\partial\omega_{2}}\;\left(\omega_{ \rm{r}  }=0\right).
\label{limitation_1}
\end{equation} 
So we can define $\omega_{-}=\frac{1}{\sqrt{\Omega_{1}^{2}+\Omega_{2}^{2}}}\left(\Omega_{2}\omega_{1}-\Omega_{1}\omega_{2}\right),\:\omega_{+}=\frac{1}{\sqrt{\Omega_{1}^{2}+\Omega_{2}^{2}}}\left(\Omega_{1}\omega_{1}+\Omega_{2}\omega_{2}\right),$
and then supplementary equation \ref{limitation_1} implies that $\forall t\;\frac{\partial H}{\partial\omega_{-}}=0\,\left(\omega_{ \rm{r}  }=0\right),$ and thus $\Delta\omega_{ \rm{r}  }\rightarrow\infty.$ 

Now consider the most general case: amplitudes and phases are not necessarily identical:
\begin{equation}
H=\underset{i}{\sum}\Omega_{i}\sin\left(\omega_{i}t+\varphi_{i}\right)\sigma_{z}.
\end{equation}
Given $\omega_{ \rm{r}  }t\ll1,$ the Hamiltonian reads:
\begin{equation}
H=\sigma_{z}\left[\underset{i}{\sum}\Omega_{i}\sin\left(\omega_{ \rm{s}  }t+\varphi_{i}\right)+\omega_{ \rm{r}  }t\underset{i}{\sum}\Omega_{i}\cos\left(\omega_{ \rm{s}  }t+\varphi_{i}\right)+\mathcal{O}\left(\omega_{ \rm{r}  }^{2}t^{2}\right)\right],
\end{equation}
so neglecting the $\left(\omega_{ \rm{r}  }t\right)^{2}$ terms the Hamiltonian can be written as:
\begin{equation}
H\approx\left[a\sin\left(\omega_{ \rm{s}  }t+\alpha\right)+b\omega_{ \rm{r}  }t\sin\left(\omega_{ \rm{s}  }t+\beta\right)\right]\sigma_{z},
\label{degeneracy}
\end{equation}
where $a\sin\left(\omega_{ \rm{s}  }t+\alpha\right)=\underset{i}{\sum}\Omega_{i}\sin\left(\omega_{ \rm{s}  }t+\varphi_{i}\right)$, and $b\sin\left(\omega_{ \rm{s}  }t+\beta\right)=\underset{i}{\sum}\Omega_{i}\cos\left(\omega_{ \rm{s}  }t+\varphi_{i}\right)$. 
It would be more convenient then to work with the parameters $\omega_{ \rm{r}  },\omega_{ \rm{s}  },a,b,\alpha,\beta$ (instead of $\omega_{1},\omega_{2},\Omega_{1},\Omega_{2},\phi_{1},\phi_{2}$).
Supplementary eq. \ref{degeneracy} immediately implies that the Hamiltonian suffers from a degeneracy, i.e. it depends only on $5$ parameters: $a,\alpha, \omega_{ \rm{s}  }, b\omega_{ \rm{r}  }, \beta.$ Namely we cannot get information on $b,$ $\omega_{ \rm{r}  }$ separately, but only on $b\omega_{ \rm{r}  }.$
A different way to phrase this is that $-\omega_{ \rm{r}  }\frac{\partial H}{\partial\omega_{ \rm{r}  }}+b\frac{\partial H}{\partial b}=0,$hence there exists a parameter $g$ such that $\forall t\;\frac{\partial H}{\partial g}=0.$
More elaborately we can see explicitly that $\text{span}\left(\left\{ \nabla f\left(t\right)\right\} _{t}\right)$ is of dimension $\leq5$ (where $H=f\left(t\right)\, \sigma_{z}$):
\begin{eqnarray}
\begin{split}
&\nabla f\left(t\right)=\left(\begin{array}{c}
\frac{\partial f}{\partial\omega_{ \rm{s}  }}\\
\frac{\partial f}{\partial\omega_{ \rm{r}  }}\\
\frac{\partial f}{\partial a}\\
\frac{\partial f}{\partial b}\\
\frac{\partial f}{\partial\alpha}\\
\frac{\partial f}{\partial\beta}
\end{array}\right)=\left(\begin{array}{c}
at\cos\left(\omega_{ \rm{s}  }t+\alpha\right)+b\omega_{ \rm{r}  }t^{2}\cos\left(\omega_{ \rm{s}  }t+\beta\right)\\
bt\sin\left(\omega_{ \rm{s}  }t+\beta\right)\\
\sin\left(\omega_{ \rm{s}  }t+\alpha\right)\\
\omega_{ \rm{r}  }t\sin\left(\omega_{ \rm{s}  }t+\beta\right)\\
a\cos\left(\omega_{ \rm{s}  }t+\alpha\right)\\
b\omega_{ \rm{r}  }t\cos\left(\omega_{ \rm{r}  }t+\beta\right)
\end{array}\right)
\\
& =t\cos\left(\omega_{ \rm{s}  }t\right)\left(\begin{array}{c}
a\cos\left(\alpha\right)\\
b\sin\left(\beta\right)\\
0\\
\omega_{ \rm{r}  }\sin\left(\beta\right)\\
0\\
b\omega_{ \rm{r}  }\cos\left(\beta\right)
\end{array}\right)+t\sin\left(\omega_{ \rm{s}  }t\right)\left(\begin{array}{c}
-a\sin\left(\alpha\right)\\
b\cos\left(\beta\right)\\
0\\
\omega_{ \rm{r}  }\cos\left(\beta\right)\\
0\\
-b\omega_{ \rm{r}  }\sin\left(\beta\right)
\end{array}\right)+\cos\left(\omega_{ \rm{s}  }t\right)\left(\begin{array}{c}
0\\
0\\
\sin\left(\alpha\right)\\
0\\
a\cos\left(\alpha\right)\\
0
\end{array}\right)+\sin\left(\omega_{ \rm{s}  }t\right)\left(\begin{array}{c}
0\\
0\\
\cos\left(\alpha\right)\\
0\\
-a\sin\left(\alpha\right)\\
\\
\end{array}\right)+b\omega_{ \rm{r}  }t^{2}\left(\begin{array}{c}
1\\
0\\
0\\
0\\
0\\
0
\end{array}\right).           
\end{split}         
\end{eqnarray}
Hence for $\omega_{ \rm{r}  }\neq0,$ the dimension is $5.$
Note that for $\omega_{ \rm{r}  }=0$ the dimension is $4$ (as $\frac{\partial H}{\partial b}=0,\frac{\partial H}{\partial\beta}=0$).
Therefore in this case as well $\Delta \omega_{ \rm{r}  } \rightarrow \infty.$

\section{Effective Hamiltonian: Conditions for a non vanishing FI}
\label{effective_Hamiltonian}

The exact expression of the accumulated phase for a single frequency signal  (given in the methods section) reads:
\begin{equation}
\phi=A\sin\left(\delta t\right)\frac{\tan\left(\omega\frac{\tau}{2}\right)}{\omega}+B\left(1-\cos\left(\delta t\right)\right)\frac{\tan\left(\omega\frac{\tau}{2}\right)}{\omega}.
\end{equation} 
Based on this expression, we want to understand what the conditions are on $\delta$ (or equivalently $\tau$) to nullify $\phi$ for every $A,B.$
Clearly, whenever $\omega \tau \neq n \pi$ and $\delta t= 2 \pi k$ ($n,k$ are integers) $\phi=0.$
Note that the condition $\delta t=2 \pi k$ implies that $\omega t=\pi m$ ($k,m$ are integers, because the total time is an integer multiple of $\frac{\pi}{\omega+\delta}$). 
In addition, whenever $\omega \tau =2 \pi n$ we have that $\tan\left(\omega\frac{\tau}{2}\right)=0$ and thus $\phi=0.$ It is simple to understand this case. The signal completes an integer number of cycles between two consecutive pulses and thus the accumulated phase vanishes.
Note that for $\omega\tau=\left(2n+1\right)\pi,$ it is not possible to nullify $\phi$ for every $A,B.$ Therefore the condition $\delta t= 2 \pi k$ is not valid here. In this case: $\phi=A\left(\frac{2}{\pi}\right)\frac{t}{2n+1},$ therefore $A\neq0\Rightarrow\phi\neq0.$
For a two frequency signal the accumulated phase is:
\begin{equation}
H=\underset{i}{\sum}\left[A_{i}\sin\left(\omega_{i}t\right)+B_{i}\cos\left(\omega_{i}t\right)\right]h\left(t\right)\sigma_{z}.
\end{equation}  
In order to get a non vanishing $I_{ \rm{r}  }$ when $\omega_{ \rm{r}  }=0$ we have to nullify $\phi$ (at $\omega_{ \rm{r}  }=0$). 
As shown above we need to either take $\delta_{s}t=2 \pi k$ or take $\omega_{ \rm{s}  } \tau= 2 \pi n.$
For the first possibility, given that $\omega_{ \rm{r}  } t \ll 1,$ the accumulated phase reads:
\begin{equation}
\phi=\underset{i}{\sum}\frac{\tan\left(\omega_{i}\frac{\tau}{2}\right)}{\omega_{i}}\left[A_{i}\sin\left(\delta_{i}t\right)+B_{i}\left(1-\cos\left(\delta_{i}t\right)\right)\right],
\end{equation}
and thus a finite $I_{ \rm{r}  }$ is obtained:
\begin{equation}
I_{ \rm{r}  }=8\sigma^{2}\frac{\tan\left(\omega_{ \rm{s}  }\frac{\tau}{2}\right)^{2}}{\omega_{ \rm{s}  }^{2}}t^{2}.
\label{generalized_fi}
\end{equation}
Note that this expression is not valid for $\omega_{ \rm{s}  }\tau=\left(2n+1\right)\pi,$ as for these values $\phi\neq0\Rightarrow I_{ \rm{r}  } = 0.$
For the second possibility ($\omega_{ \rm{s}  } \tau= 2 \pi n$): Note that this option has an overlap with the first one, in that if $\frac{t}{\tau}$ is even then $\delta t$ is an integer multiplication of $2\pi.$ However it can be seen that in this case $\phi\sim\omega_{ \rm{r}  }^{2}\Rightarrow p\sim\omega_{ \rm{r}  }^{4}$ and thus $I_{ \rm{r}  }$ vanishes.
If $\frac{t}{\tau}$ is odd (and $\omega_{ \rm{s}  } \tau= 2 \pi n$) then: $\phi\approx\frac{\left(B_{1}-B_{2}\right)\tau^{2}}{2\pi n}\omega_{ \rm{r}  }\Rightarrow I_{ \rm{r}  }=\frac{8\sigma^{2}\tau^{4}}{\left(2\pi\right)^{2}n^{2}}.$ Hence a finite $I_{ \rm{r}  }$ is achieved but it is much lower than with the first option, as it independent of $t,$ and thus does not grow with $t.$
Therefore we dismiss this option and only keep the first one, in which $\delta_{s}t=2\pi k\;\left(\omega_{ \rm{s}  }\tau\neq n\pi\right),$ and the FI is given in supplementary equation \ref{generalized_fi}. 

Naturally we would like to confirm that the optimal detuning (or $\tau$) is $\delta_{s}= \pm 2 \pi/t.$ 
To see this, let us first examine how close $\omega_{ \rm{s}  } \tau$ can approach $\pi$ (while requiring $\delta_{s}t=2\pi k\neq0$):
\begin{equation}
\omega\tau=\left(\omega+\delta\right)\tau-\delta\tau=\pi-\delta\tau.
\end{equation}
Hence the closest it can approach $\pi$ is by $\delta \tau.$ Since the minimal possible $\delta$ is $\frac{2 \pi}{t}$,we cannot get closer to $\pi$ than $\frac{2 \pi \tau}{t}=\frac{2 \pi}{N}$
(where $N$ is the number of pulses). 
Therefore if the closest we can get to $\pi$ is $\frac{2 \pi}{N}\approx\frac{2\pi^{2}}{\omega t},$ then the closest we can get to $3\pi$ is $\approx\frac{6\pi^{2}}{\omega t},$ and so on.
Therefore the optimal $I_{ \rm{r}  }$ is achieved with $\delta_{s}=\pm\frac{2\pi}{t}$ and reads:     
\begin{equation}
I_{ \rm{r}  }=8\sigma^{2}\frac{\tan\left(\frac{\pi}{2\left(1\pm\frac{2\pi}{\omega_{ \rm{s}  }t}\right)}\right)^{2}}{\omega_{ \rm{s}  }^{2}}t^{2}\approx\frac{8}{\pi^{4}}\sigma^{2}t^{4}.
\end{equation}
It is well established that for frequency estimation problems the optimal scaling of the FI is $\sim \Omega^{2}t^{4}$\cite{pang2017optimal, schmitt2017submillihertz, jordan2017classical, yang2017quantum, gefen2017control, naghiloo2017achievingT4}, where $\Omega$ stands for the amplitude of the signal, and this is exactly the scaling that this method achieves. The optimality of this method is discussed in supplementary note \ref{optimality}.

\section{Optimality analysis}
\label{optimality}
We presented control methods for which $I_{ \rm{r}  }\neq0$ , and found the optimal one out of a set of possible controls.
However a valid question is whether this method is optimal out of all possible control strategies.
This question is left open; however, we can find an upper bound of $I_{ \rm{r}  }$ (which is not tight).
Given a Hamiltonian:            
\begin{equation}
H=\underset{i}{\sum}\left(A_{i}\cos\left(\omega_{i}t\right)+B_{i}\sin\left(\omega_{i}t\right)\right)\sigma_{Z}
\end{equation}
The optimal $I_{ \rm{r}  }$ (FI about $\omega_{ \rm{r}  }$) is given by (according to supplementary ref. \cite{pang2017optimal}): 
\begin{equation}
I_{ \rm{r}  }=4\left[\int \left\vert \frac{\partial H}{\partial\omega_{ \rm{r}  }} \right\vert \:dt\right]^{2},
\end{equation}
where $|\bullet|$ stands for the operator norm (in general it is the maximal eigenvalue minus the minimal; for $H\propto\sigma_{\theta}$ it can be written in this way). For $\omega_{ \rm{r}  }=0$:
\begin{equation}
\frac{\partial H}{\partial\omega_{ \rm{r}  }}=\left[-\left(A_{1}-A_{2}\right)t\sin\left(\omega_{ \rm{s}  }t\right)+\left(B_{1}-B_{2}\right)t\cos\left(\omega_{ \rm{s}  }t\right)\right]\sigma_{Z}.
\end{equation}
Therefore given that $\omega_{ \rm{s}  }t\gg1$, the maximal $I_{ \rm{r}  }$ reads:
\begin{equation}
I_{ \rm{r}  }=\left(\frac{2}{\pi}\right)^{2}\left[\left(A_{1}-A_{2}\right)^{2}+\left(B_{1}-B_{2}\right)^{2}\right] t^{4}.
\end{equation}
 Therefore given that $A_{i},B_{i}$ are i.i.d. with variance $\sigma^{2}$
an upper bound for the average $I_{ \rm{r}  }$ is: 
\begin{equation}
I_{ \rm{r}  }\leq\frac{16}{\pi^{2}}\sigma^{2}t^{4}.
\end{equation}
 Given $A_{i},B_{i}$ the control that achieves the optimal $I_{ \rm{r}  }$
consists of applying $\pi$-pulses whenever $\frac{\partial H}{\partial\omega_{ \rm{r}  }}$
flips a sign; therefore it requires knowing $A_{i},B_{i}$, which
is unrealistic in the setting described in the paper. In practice
we need to apply the same control to every realization of $A_{i},B_{i}$,
hence this upper bound is not achievable. Using the method presented
in the paper we obtain $I_{ \rm{r}  }=\frac{8\sigma^{2}t^{4}}{\pi^{4}}$ , hence lower
by a factor of $2\pi^{2}$ from this upper bound. Whether our method
is optimal given this noise model is left as an open question.


\section{Multiparameter estimation of all the parameters}
\label{sec:multiparameter_estimation}
In an actual experimental scenario all the parameters are unknown, and since the estimation protocol of $\omega_{ \rm{r}  }$
depends on $\omega_{ \rm{s}  }$ (the pulses frequency should be detuned from $\omega_{ \rm{s}  }$ by $\frac{2\pi}{\omega_{ \rm{s}  }}$), a preliminary estimation of $\omega_{ \rm{s}  }$
is necessary. 
 We propose the following protocol: 
For a preliminary estimation of all the parameters ($\omega_{ \rm{r}  },\omega_{ \rm{s}  },\sigma$) the traditional protocol is applied; namely vary the pulses frequency (denoted as $\omega_{p}$) and make a large number of measurements for each $\omega_{p}.$
The next step is to fit the obtained data points (or make MLE analysis). This should provide a good estimation of $\omega_{ \rm{s}  }$ and $\sigma,$ the estimation of $\omega_{ \rm{r}  }$ however will not be good enough (unless by chance we hit very close to the resonance points, and perform enough measurements in this point). 
The performance  of this estimation method is illustrated in supplementary fig. \ref{preliminary_estimation}.

 \begin{figure}[h]
\begin{center}
\begin{subfigure}[]
{\includegraphics[width=7.2 cm,height=3.8cm]{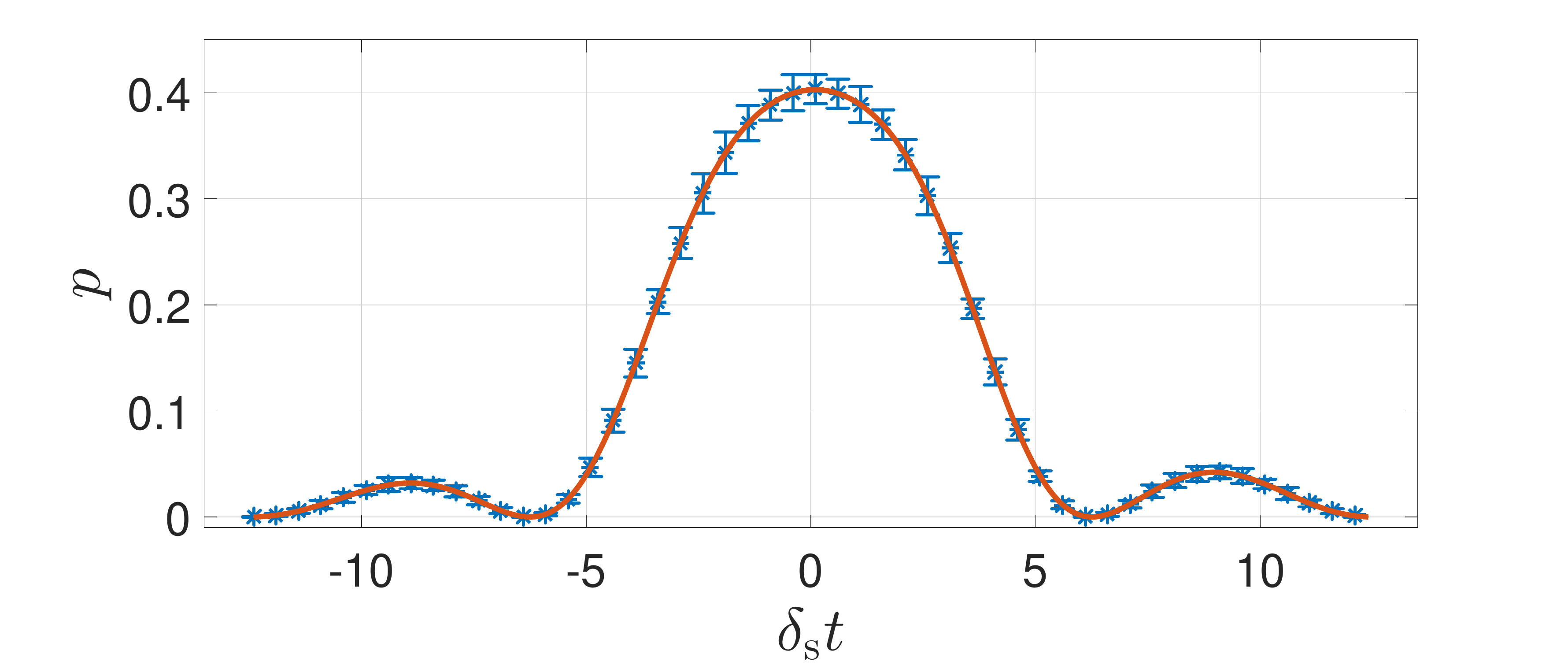}}
\end{subfigure}
\vskip 1mm
\begin{subfigure} []
\centering
{\includegraphics[width=5 cm,height=3.4cm]{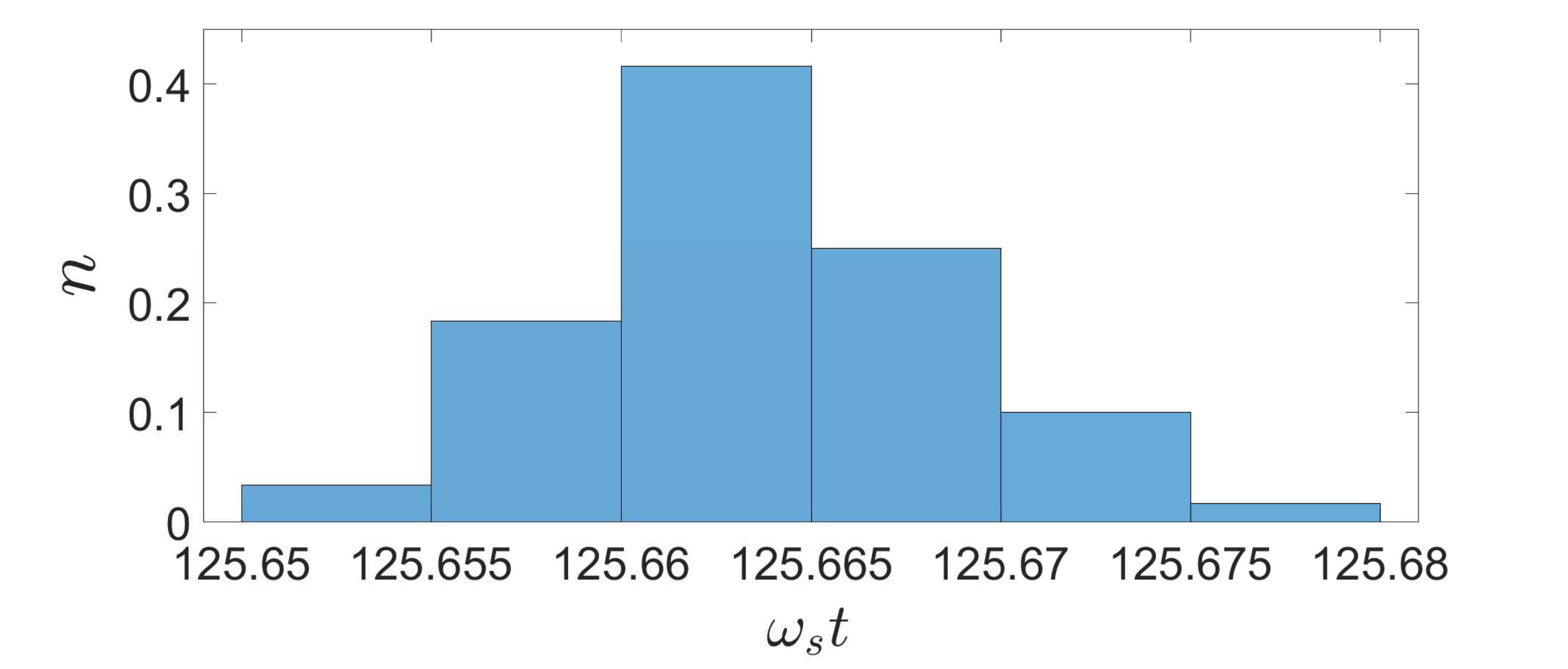}}
{\includegraphics[width=5 cm,height=3.4cm]{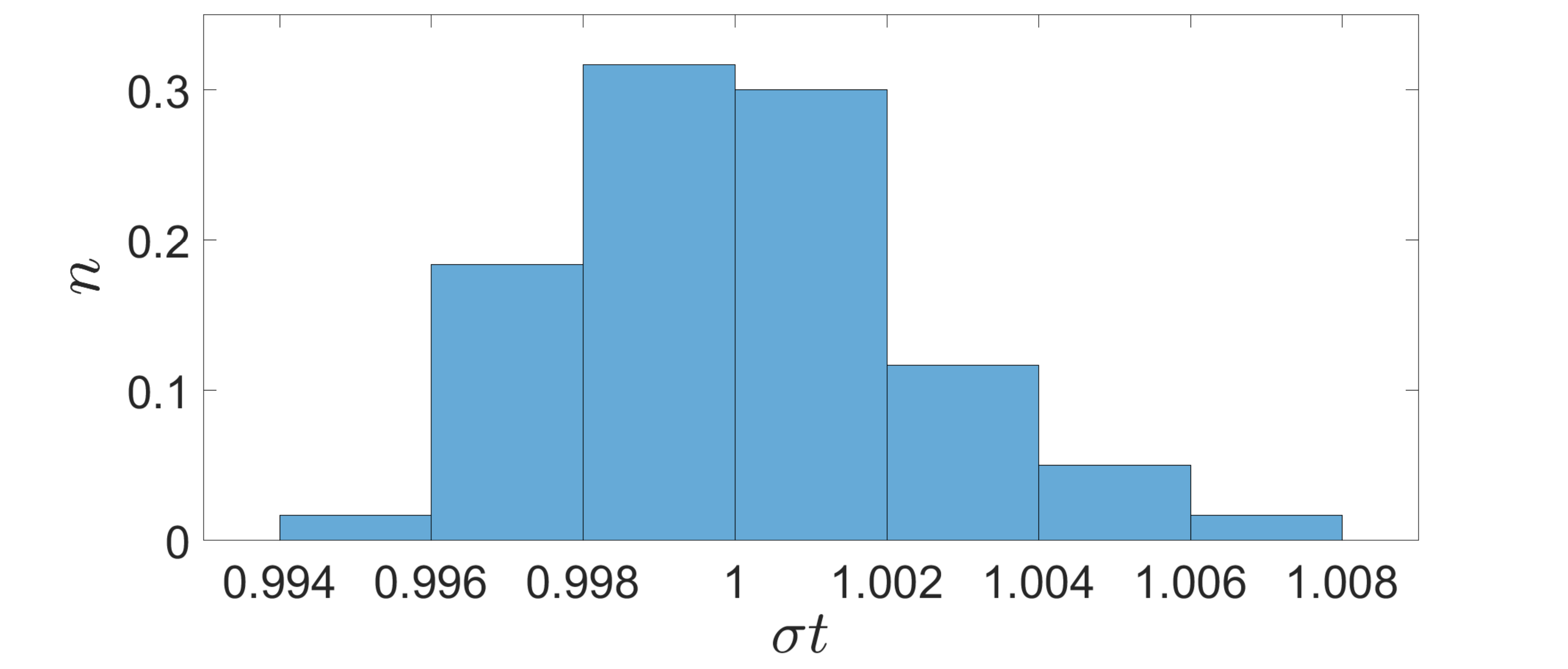}}
{\includegraphics[width=5 cm,height=3.3cm]{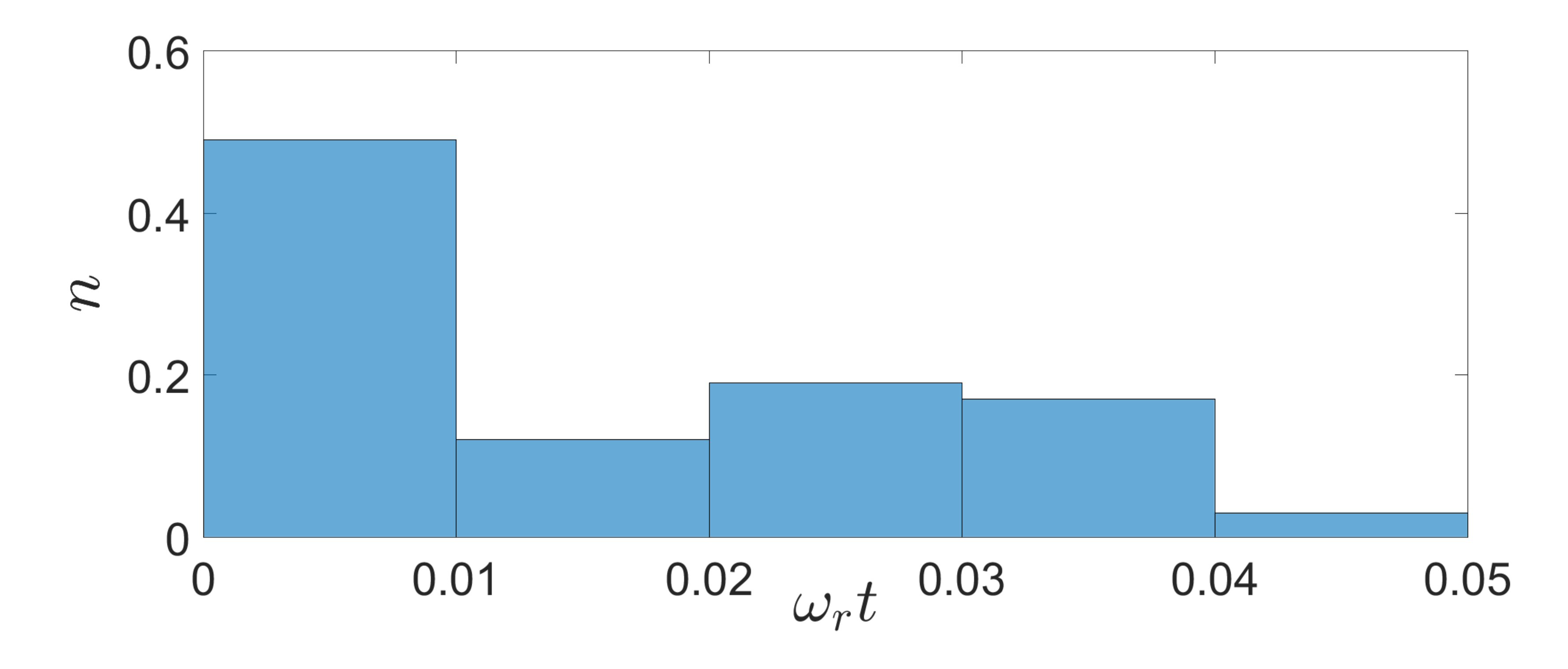}}
\end{subfigure}
\end{center}
\caption{ Preliminary estimation of the parameters achieved by varying $\delta_{s}$ and fitting the probability as a function of $\delta_{s}.$ A numerical example is presented.
The fit is shown in (a) and the estimation errors are presented in (b). While the $\omega_{ \rm{s}  }$ can be estimated efficiently , the estimation of $\omega_{ \rm{r}  }$ is not good enough for resolution.
For this illustration: $\sigma t=1, \, \omega_{ \rm{s}  }t=40 \pi $ and the number of measurements in each detuning is $n=2\cdot 10^4.$  }
\label{preliminary_estimation}
\end{figure}

The next step is to use the estimated $\omega_{ \rm{s}  },\sigma$ to apply our method. One can treat $\omega_{ \rm{s}  },\sigma$ as known and estimate $\omega_{ \rm{r}  },$ however this will create a bias (we also want to progressively improve the estimation of $\omega_{ \rm{s}  },$ such that the detuning of the pulses will be more accurate).
Note that we cannot get information about three different parameters by measuring copies of the same density matrix (even if measuring different observables, the FI matrix will be singular).
Therefore at least three different measurements are required: one measurement with $\delta_{s}t=2\pi,$ the resonance condition for estimation of $\omega_{ \rm{r}  }$ and two other measurements with the optimal detunings for estimating $\omega_{ \rm{s}  },\sigma.$ The FI about $\omega_{ \rm{s}  },\sigma$ as a function of $\delta_{s}$ is shown in supplementary fig. \ref{fi_other_parameters}.
Quite interestingly for both $\omega_{ \rm{s}  },\sigma$ the optimal $\delta_{s}\rightarrow\frac{2\pi}{t}$ as $\sigma\rightarrow\infty$ (this is because as $\sigma$ becomes larger the exponential decay becomes stronger and one needs to get closer to $\delta_{s} t=2 \pi$).

The optimal FI about $\omega_{ \rm{s}  }$ scales as $\sigma^{2}t^{4}$ and is comparable to the optimal FI about $\omega_{ \rm{r}  }.$
The optimal FI about $\sigma$ behaves in an unusual manner: Usually the FI about the amplitude grows as $t^2,$ while here the optimum (for $\sigma t >1$) is $I_{\sigma}\sim\frac{0.63}{\sigma^{2}}.$ It does not depend on $t,$ and it drops as $\sigma$ gets larger.
This behavior is somewhat similar to sensing the standard deviation of the amplitude of a stationary signal ($H=A\sigma_{Z}$ where $A\sim N\left(0,\sigma\right)$). 

Since we are dealing with a multivariable estimation, the Cram{\'e}r-Rao bound is given by the Fisher information matrix in the following way \cite{cramer2016mathematical}: $\left(\Delta x\right)^{2}=\left(I^{-1}\right)_{x,x}.$
We would like then to calculate the FI matrix. The full expression of the matrix is quite involved,
however we can easily observe that as $\omega_{\rm{r}}  \rightarrow 0 $ the FI matrix converges to a block diagonal matrix, with $I_{\rm{r}}$ as one of its eigenvalues.
The FI matrix per three measurements is the sum of the FI matrices of each measurement: $I^{\left(1\right)}+I^{\left(2\right)}+I^{\left(3\right)}$.
Therefore the FI matrix per single measurement reads: $I=\frac{1}{3}\left(I^{\left(1\right)}+I^{\left(2\right)}+I^{\left(3\right)}\right).$
Denoting the FI matrix that corresponds to $\delta_{\rm{s}} t= 2 \pi$ as $I^{\left(1\right)}$, we observe that:
\begin{equation}
\frac{\left(\frac{\partial p}{\partial\omega_{{\rm s}}}\right)^{2}}{p\left(1-p\right)},\frac{\left(\frac{\partial p}{\partial\sigma}\right)^{2}}{p\left(1-p\right)},\frac{\left(\frac{\partial p}{\partial\sigma}\right)\left(\frac{\partial p}{\partial\omega_{{\rm r}}}\right)}{p\left(1-p\right)},\frac{\left(\frac{\partial p}{\partial\omega_{{\rm s}}}\right)\left(\frac{\partial p}{\partial\omega_{{\rm r}}}\right)}{p\left(1-p\right)}\rightarrow0\Longrightarrow I^{\left(1\right)}=\left(\begin{array}{ccc}
I_{{\rm r}} & 0 & 0\\
0 & 0 & 0\\
0 & 0 & 0
\end{array}\right)  
\end{equation}
Regarding $I^{\left(2\right)},I^{\left(3\right)}$, note that for them (since $\delta_{s}t\neq2\pi n$) we have that:
\begin{equation}
\frac{\left(\frac{\partial p}{\partial\omega_{{\rm r}}}\right)^{2}}{p\left(1-p\right)},\frac{\left(\frac{\partial p}{\partial\omega_{{\rm r}}}\right)\left(\frac{\partial p}{\partial\omega_{{\rm s}}}\right)}{p\left(1-p\right)},\frac{\left(\frac{\partial p}{\partial\omega_{{\rm r}}}\right)\left(\frac{\partial p}{\partial\sigma}\right)}{p\left(1-p\right)}\rightarrow0\Longrightarrow I^{\left(j\right)}=\left(\begin{array}{ccc}
0 & 0 & 0\\
0 & I_{2,2}^{\left(j\right)} & I_{2,3}^{\left(j\right)}\\
0 & I_{3,2}^{\left(j\right)} & I_{3,3}^{\left(j\right)}
\end{array}\right).
\end{equation}     
Therefore the FI matrix per single measurement reads:
\begin{equation}
I=\frac{1}{3} \left(\begin{array}{ccc}
I_{{\rm r}} & 0 & 0\\
0 & I_{2,2} & I_{2,3}\\
0 & I_{3,2} & I_{3,3}
\end{array}\right),
\end{equation}                  
and thus $\Delta \omega_{ \rm{r}  }=\sqrt{  \frac{ 3  }{ I_{r} } }$, we therefore get an extra factor of $\sqrt{3}$ due to these chunks of three measurements.   
The standard deviation obtained in practice (with MLE) is a bit above the expected analytical values, as can be seen in supplementary fig. \ref{multiparameter_estimation}.    
    
 \begin{figure}[h]
\begin{center}
\subfigure[]{\includegraphics[width=5.7 cm]{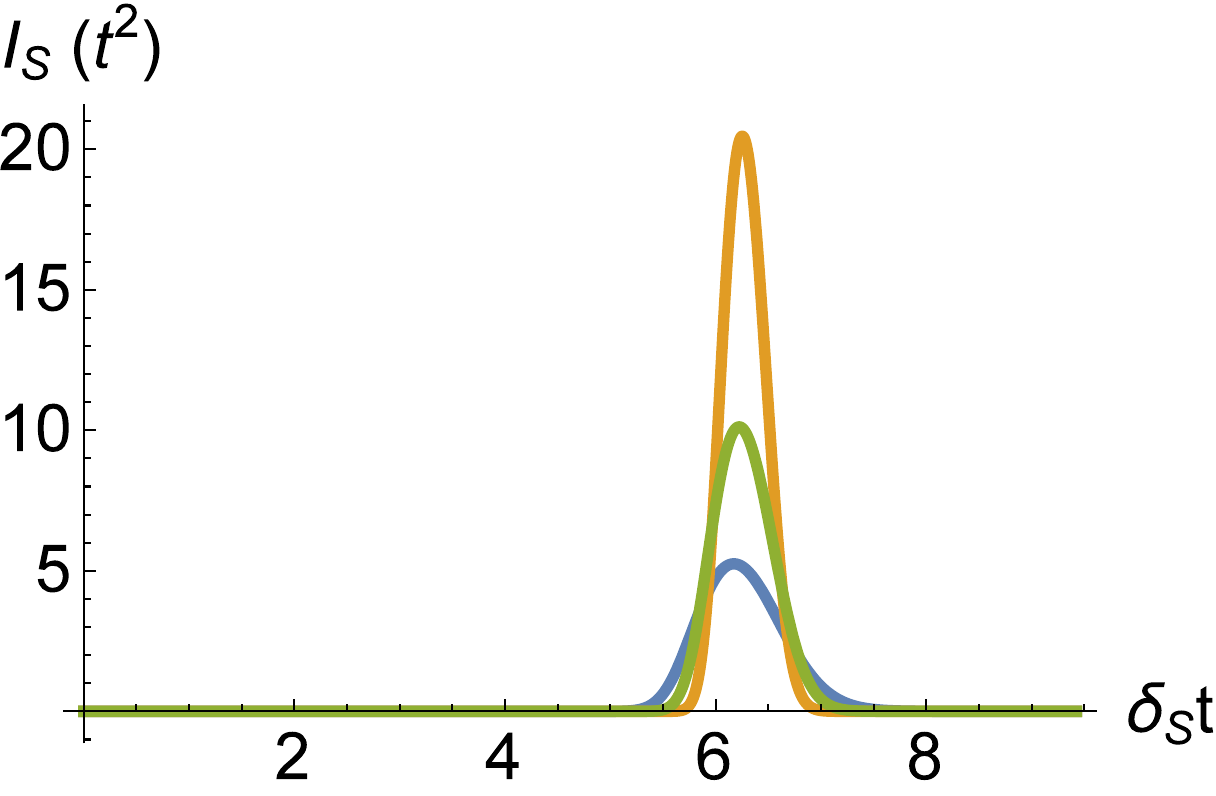}}
\subfigure[]{\includegraphics[width=5.7 cm]{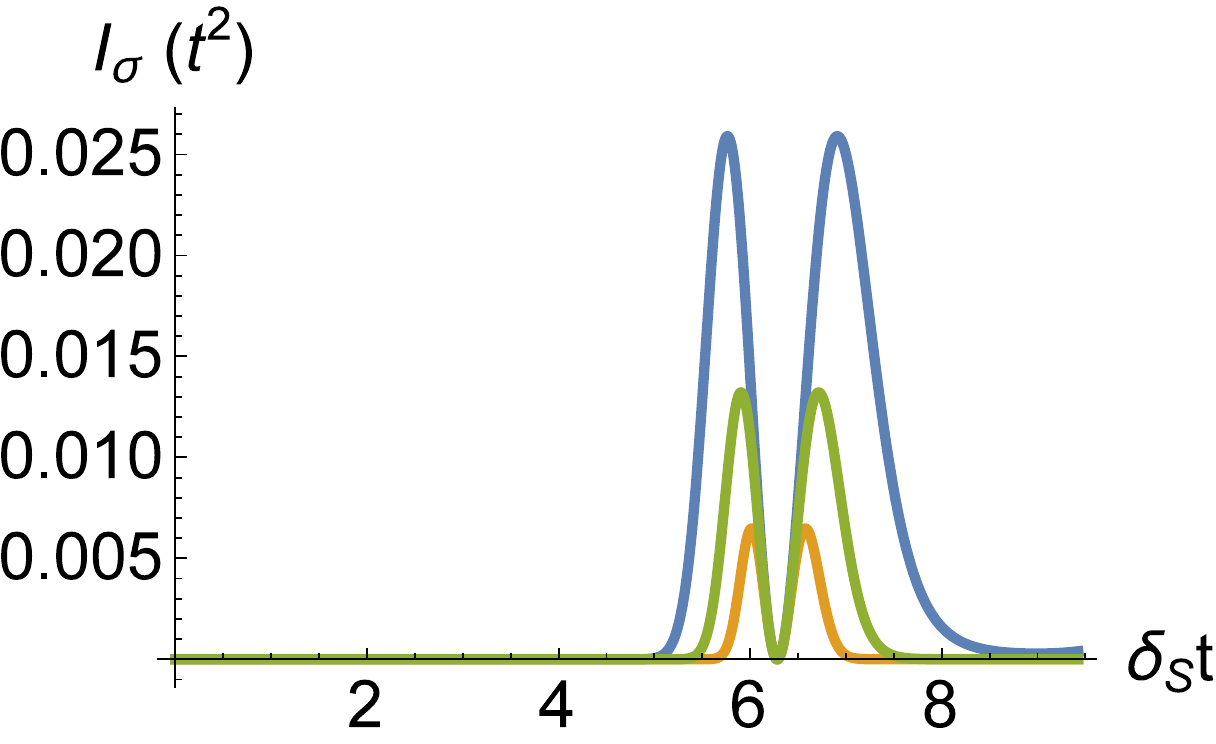}}
\end{center}
\caption{ (a) The FI about $\omega_{ \rm{s}  }$ as function of $\delta_{s}t$ for different values of $\sigma$ ($\sigma t=5$:\, Orange,  $\sigma t=7$:\, Green, $\sigma t=10$:\, Blue ).\\
(b) The FI about $\sigma$ as function of $\delta_{s}t$ for different values of $\sigma$ (same colors).
In both plots $\omega_{ \rm{r}  }=0.$
Interestingly, as $\sigma t\rightarrow\infty$ the optimal $\delta_{s} t \rightarrow 2\pi.$ 
  }
\label{fi_other_parameters}
\end{figure}

\begin{figure}[h]
\begin{center}
\begin{subfigure} []
\centering
{\includegraphics[width=5 cm,height=3.2cm]{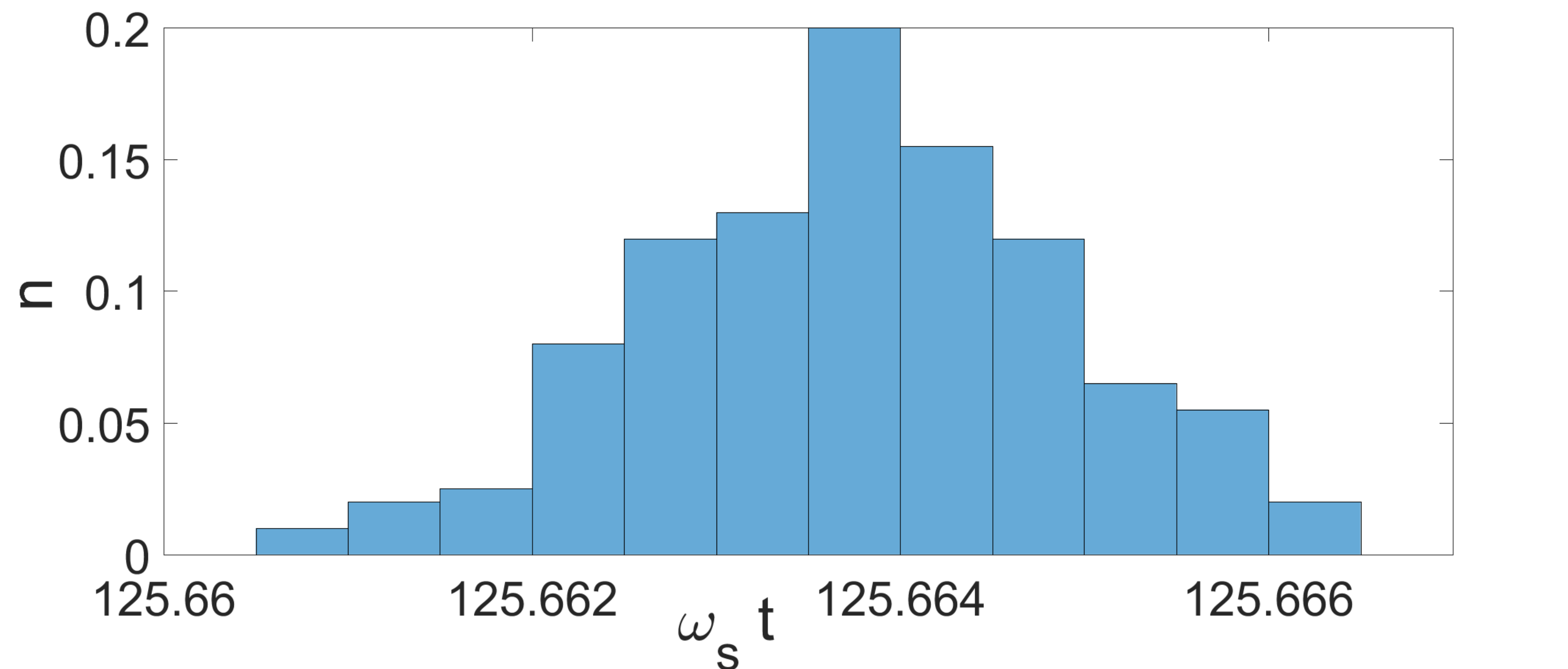}}
{\includegraphics[width=5 cm,height=3.4cm]{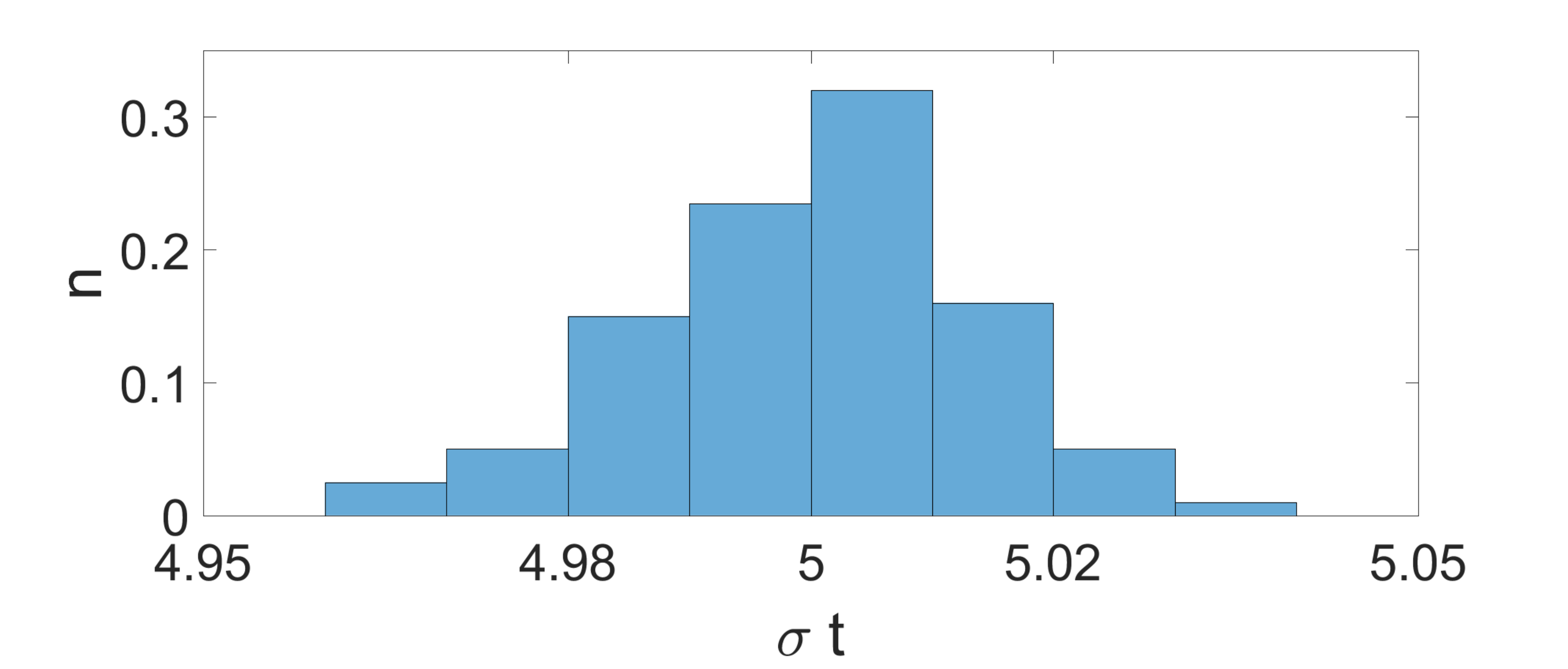}}
{\includegraphics[width=5 cm,height=3.3cm]{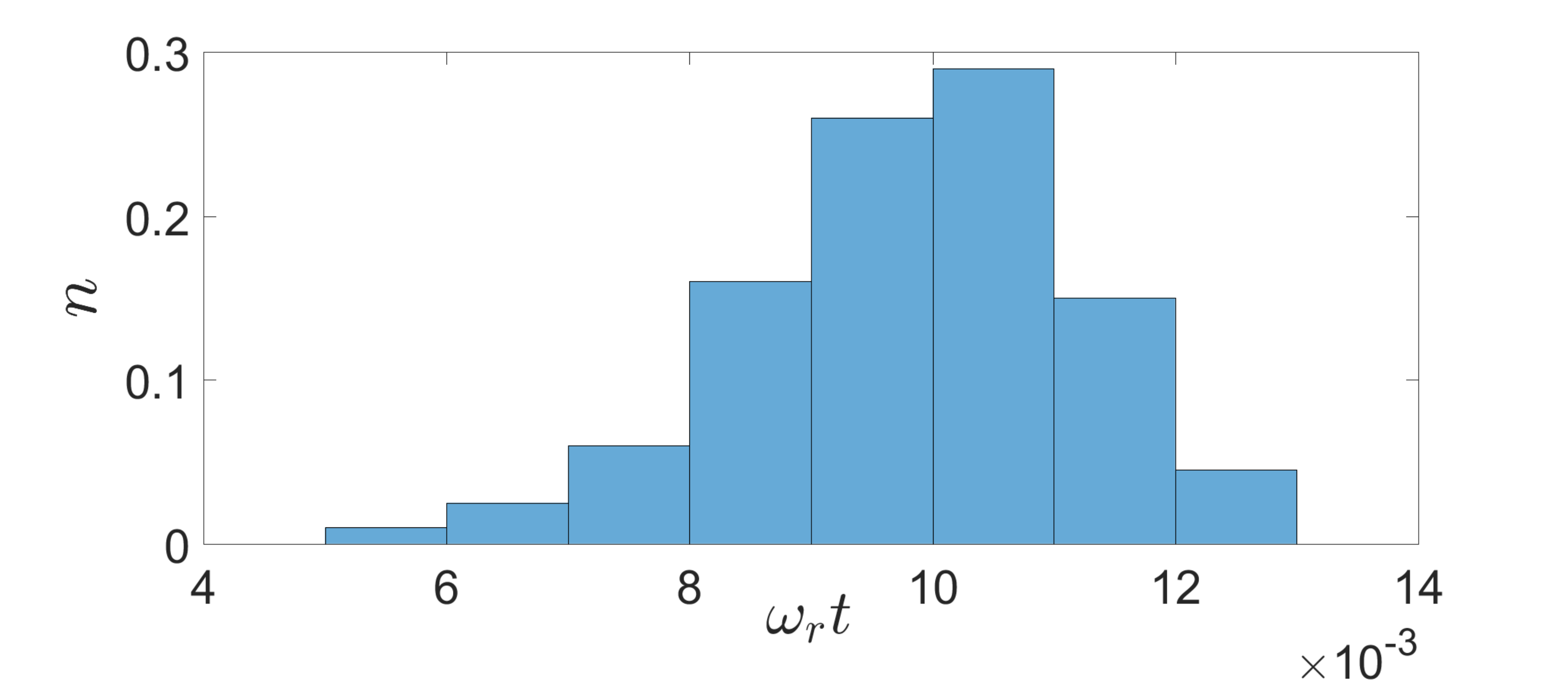}}
\end{subfigure}
\vskip 1mm
\begin{subfigure}[]
{\includegraphics[width=9 cm]{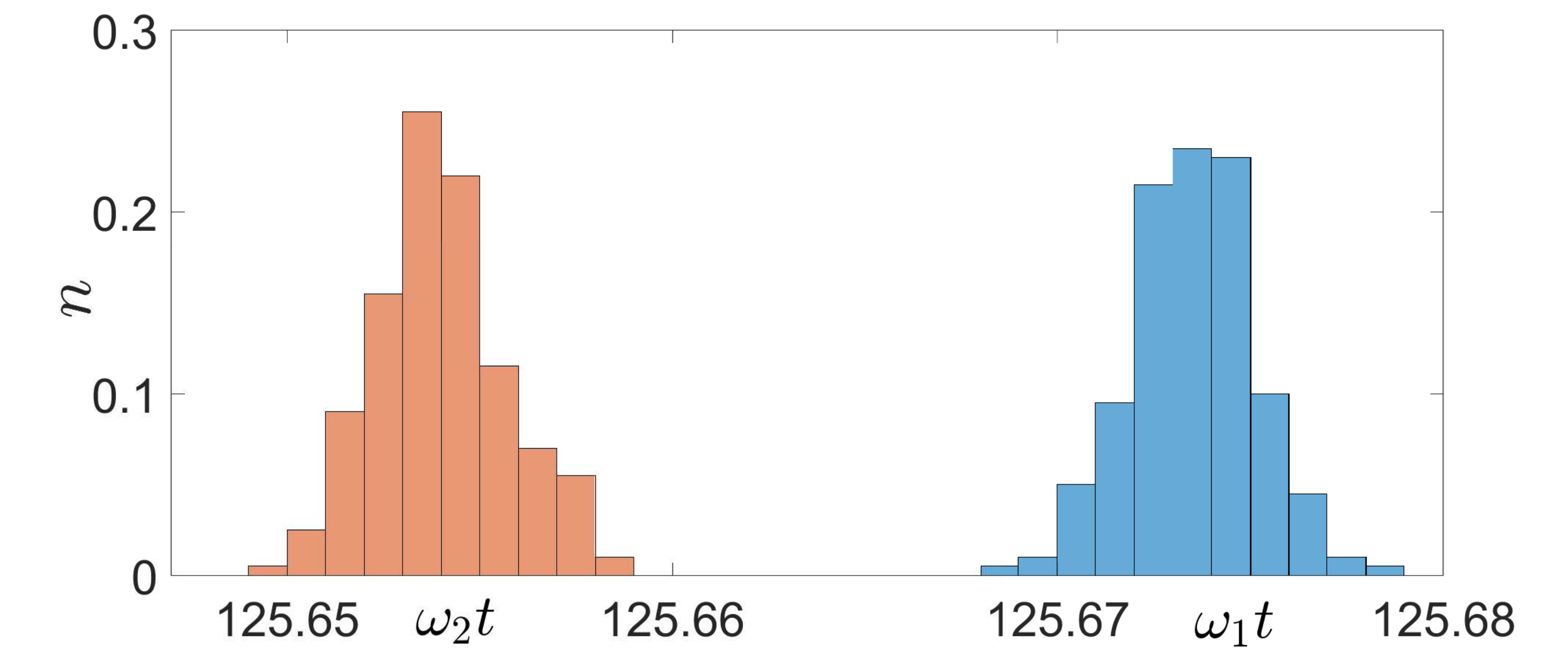}}
\end{subfigure}
\vskip 1mm
\begin{subfigure}[]
{\includegraphics[width=9 cm]{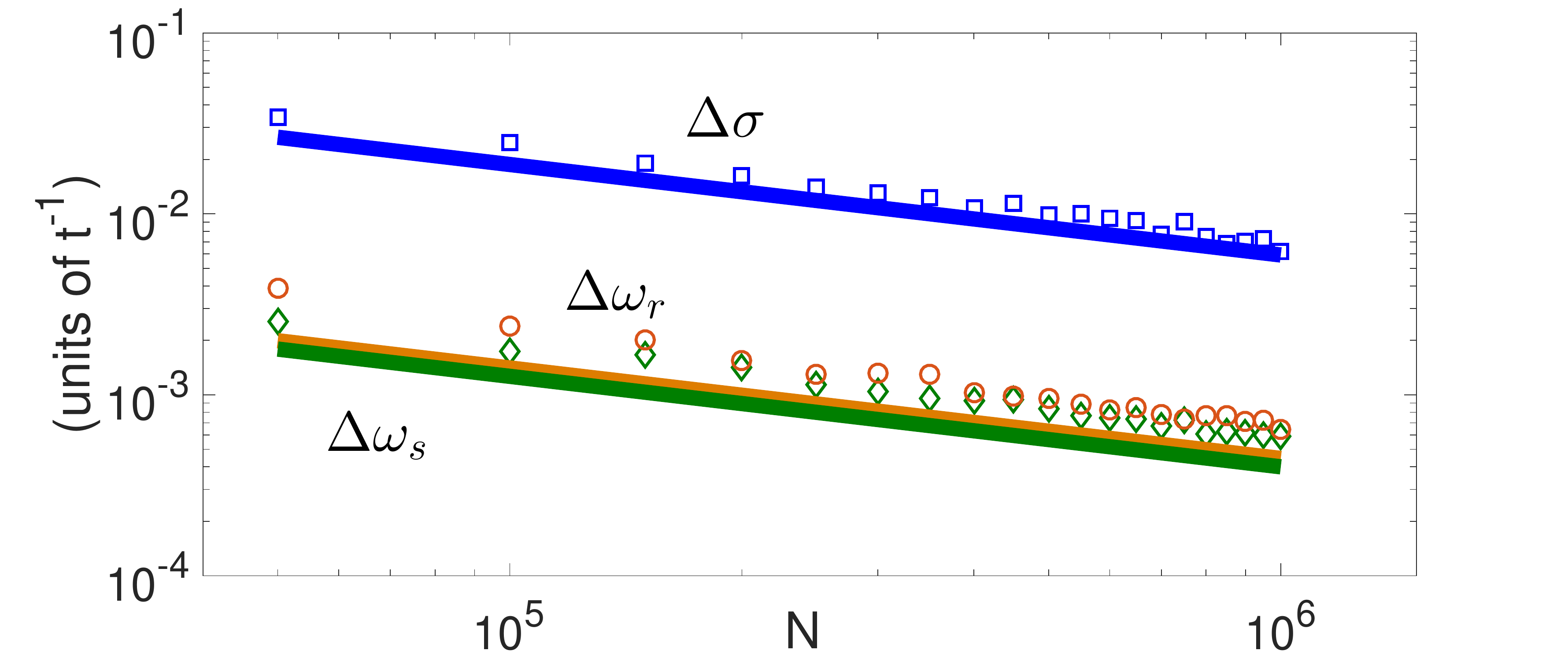}}
\end{subfigure}
\end{center}
\caption{The second stage of the protocol consists of measurements in three different detunings (each one is optimal for the estimation of a different parameter; one of them is therefore $\delta_{s}=\frac{2\pi}{t}$).
In this example: $\delta_{s}t=0.01,$ $\sigma t=5$ and the number of measurements in each detuning is $N=3\cdot 10^{5}.$
The histograms for the different parameters are presented in (a). It can be seen that resolution is achieved as $\Delta\omega_{ \rm{r}  }=\frac{1}{7.5}\omega_{ \rm{r}  }.$
In (b) the histograms of $\omega_{1},\omega_{2}$ are presented, it can be seen that the frequencies are clearly resolved.  
(c) The RMSE (root mean square error) in estimating the parameters as a function of $N.$ The solid lines are the theoretical expectations (green (bottom): $\Delta \omega_{ \rm{s}  },$ orange (middle): $\Delta \omega_{ \rm{r}  },$ blue (top): $\Delta \sigma$). 
The points represent the RMSE achieved in an actual maximum likelihood estimation (green (diamonds):$\Delta \omega_{ \rm{s}  },$ orange (circles): $\Delta \omega_{ \rm{r}  },$ blue (squares): $\Delta \sigma$).
It can be seen that there is no divergence: The RMSE of all the parameters scale as $N^{-0.5}.$ In fact $\Delta \omega_{ \rm{r}  }$ is very close to $\Delta \omega_{ \rm{s}  },$ whereas the worst is $\Delta \sigma$.  }
\label{multiparameter_estimation}
\end{figure}  

\section{Limitation due to incoherence}
\label{sec:incoherence}
In this part we derive the effect of incoherence of the signal (during the measurement) on the method, where we consider a model in which the quadratures undergo identical and independent OU process.
The OU process is defined as $dA_{i}=-\gamma A_{i}\,dt+\sigma_{n}\,dW_{t}^{A_{i}},$ and similarly for $B_{i}.$
To get the effect, we need to calculate the transition probability $p=\langle\sin\left(\phi\right)^{2}\rangle ,$ where $\phi$ is the accumulated phase. Note that:
\begin{equation}
\phi=\overset{2}{\underset{i=1}{\sum}}\overset{T}{\underset{0}{\int}}\left(A_{i}\left(t\right)\cos\left(\omega_{i}t\right)+B_{i}\left(t\right)\sin\left(\omega_{i}t\right)\right)\:dt,
\end{equation}
where $A_{i}\left(t\right)=A_{i}\left(0\right)\exp\left(-\gamma t\right)+\sigma_{n}\overset{t}{\underset{0}{\int}}e^{-\gamma\left(t-s\right)}dW_{s}^{A_{i}},$ and the same holds for $B_{i}\left(t\right).$ Therefore $\phi$ can be written as the sum $\phi=\phi_{av}+\phi_{n},$ where:
\begin{equation} 
\phi_{av}=\overset{2}{\underset{i=1}{\sum}}\overset{T}{\underset{0}{\int}}\left(A_{i}\left(0\right)e^{-\gamma t}\cos\left(\omega_{i}t\right)+B_{i}\left(0\right)e^{-\gamma t}\sin\left(\omega_{i}t\right)\right)\:dt,
\end{equation}
and:
\begin{equation}
\phi_{n}=\overset{2}{\underset{i=1}{\sum}}\sigma_{n}\overset{T}{\underset{0}{\int}}dt\:\left(\cos\left(\omega_{i}t\right)\underset{0}{\overset{t}{\int}}e^{-\gamma\left(t-s\right)}dW_{s}^{A_{i}}+\sin\left(\omega_{i}t\right)\underset{0}{\overset{t}{\int}}e^{-\gamma\left(t-s\right)}dW_{s}^{B_{i}}\right).
\end{equation}
Observe that $\phi_{n}$ is a Gaussian random variable with $\langle\phi_{n}\rangle=0,$ and:
\begin{equation}
\langle\phi_{n}^{2}\rangle=\overset{2}{\underset{i=1}{\sum}}\sigma_{n}^{2}\overset{T}{\underset{0}{\int}}ds\left[\underset{s}{\overset{T}{\int}}\cos\left(\omega_{i}t\right)e^{-\gamma\left(t-s\right)}\:dt\right]^{2}+\underset{i}{\sum}\sigma_{n}^{2}\overset{T}{\underset{0}{\int}}ds\left[\underset{s}{\overset{T}{\int}}\sin\left(\omega_{i}t\right)e^{-\gamma\left(t-s\right)}\:dt\right]^{2}
\end{equation}
 We would like now to take $\omega_{ \rm{s}  }T=2\pi$ in the regime of: $\gamma T\ll1,\:\phi_{n}\ll1,\:\omega_{ \rm{r}  }T\ll1.$ In this case $p=\langle\sin\left(\phi\right)^{2}\rangle\approx\langle\phi_{av}^{2}\rangle+\langle\phi_{n}^{2}\rangle.$ In leading orders:
\begin{equation}
\langle\phi_{n}^{2}\rangle\approx\sigma_{n}^{2}T^{3}\left(\frac{1}{\pi^{2}}+O\left(\gamma T\right)\right),
\end{equation}
and:
\begin{eqnarray}
\begin{split}
&\langle\phi_{av}^{2}\rangle\approx\langle\left[\frac{\left(A_{1}\left(0\right)-A_{2}\left(0\right)\right)}{2\pi}\omega_{ \rm{r}  }T^{2}+\overset{2}{\underset{i=1}{\sum}}B_{i}\left(0\right)\frac{\gamma T^{2}}{2\pi}\right]^{2}\rangle=\\
&=\overset{2}{\underset{i=1}{\sum}}\langle A_{i}\left(0\right)^{2}\rangle\frac{1}{4\pi^{2}}\omega_{ \rm{r}  }^{2}T^{4}+\overset{2}{\underset{i=1}{\sum}}\langle B_{i}\left(0\right)^{2}\rangle\frac{1}{4\pi^{2}}\gamma^{2}T^{4}=\\
&=\frac{\sigma_{n}^{2}}{4\pi^{2}\gamma}\omega_{ \rm{r}  }^{2}T^{4}+\frac{\sigma_{n}^{2}}{4\pi^{2}}\gamma T^{4}.
\end{split}
\end{eqnarray}
The second term ($\underset{i}{\sum}\frac{\sigma_{n}^{2}}{4\pi^{2}}\gamma T^{4}$) can be neglected as it is much smaller than $\langle\phi_{n}^{2}\rangle\sim\sigma_{n}^{2}T^{3}.$ Therefore: 
\begin{equation}
p\approx\langle\phi_{av}^{2}\rangle+\langle\phi_{n}^{2}\rangle\approx\frac{\sigma_{n}^{2}T^{3}}{\pi^{2}}+\frac{\sigma_{n}^{2}}{4\pi^{2}\gamma}\omega_{ \rm{r}  }^{2}T^{4},
\end{equation}
using the notation in the main text $\frac{\sigma_{n}^{2}T^{3}}{\pi^{2}}=n$, and the resolution condition is thus $\omega_{ \rm{r}  }^{2}\frac{T}{\gamma}\gg1.$ 

\section{Limitation due to dephasing of the probe}
\label{sec:dephasing}
Let us now find the implications of noise inflicted on the probe: specifically we consider a Markovian dephasing. Intuitively this should set an additional limitation:
the transition probability does not vanish now due to two reasons: the finite $\omega_{ \rm{r}  },$ namely the term $\left(\sigma t\right)^{2}\left(\omega_{ \rm{r}  }t\right)^{2},$ and also due to a dephasing rate of $\kappa,$ namely a term of $\kappa t.$
Hence resolution can be achieved if the first term is larger than the second, hence the condition is $\frac { \omega_{ \rm{r}  }\sigma}{\left(\kappa^{2}\right)}\gg1.$
  
In more detail, given our Hamiltonian $H=f\left(t\right)\sigma_{z}$ and a dephasing rate $\kappa,$ the time evolution is given by the Master equation:
\begin{equation}
\frac{d\rho}{dt}=-i\left[f\left(t\right)\sigma_{z},\rho\right]+\kappa\left(\sigma_{z}\rho\sigma_{z}-\rho\right).
\end{equation}
Initializing and measuring in $\sigma_{x}$ basis, we get the transition probability $p=0.5-0.5e^{-2\kappa t}\cos\left(2\phi\right)$ (where $\phi$ is the accumulated phase).
Averaging over the different realizations we get: $p=0.5\left(1-\exp\left(-\frac{\left(\sigma t\right)^{2}\left(\omega_{ \rm{r}  }t\right)^{2}}{\pi^{2}}-2\kappa t\right)\right),$ and the FI reads:
\begin{equation}
I_{ \rm{r}  }=\frac{4\omega_{ \rm{r}  }^{2}\sigma^{4}t^{8}}{\pi^{4}\left[\left(\exp\left(4\kappa t+2\omega_{ \rm{r}  }^{2}\sigma^{2}t^{4}/\pi^{2}\right)-1\right)\right]}.
\end{equation}
Hence the noiseless FI is retrieved only for $ \frac { \omega_{ \rm{r}  }\sigma}{\left(\kappa^{2}\right)}\gg1.$
 Note that the effect of dephasing in this problem is quite different: first, the FI is not necessarily close to the noiseless FI for $\kappa t \ll 1,$ the condition $ \frac { \omega_{ \rm{r}  }\sigma}{\left(\kappa^{2}\right)}\gg1$ must hold.  
 Second, since there is a competition between $\left(\sigma t\right)^{2}\left(\omega_{ \rm{r}  }t\right)^{2}$ and $\kappa t,$ and for short enough times the second term is always larger, then the noiseless FI is retrieved only after a minimal time (goes as $\sim \frac{\kappa^{1/3}}{\sigma^{2/3}\omega_{ \rm{r}  }^{2/3}}$) . This behavior is illustrated in supplementary fig. \ref{dephasing}.
 
 \begin{figure}[h]
\begin{center}
{\includegraphics[width=7 cm]{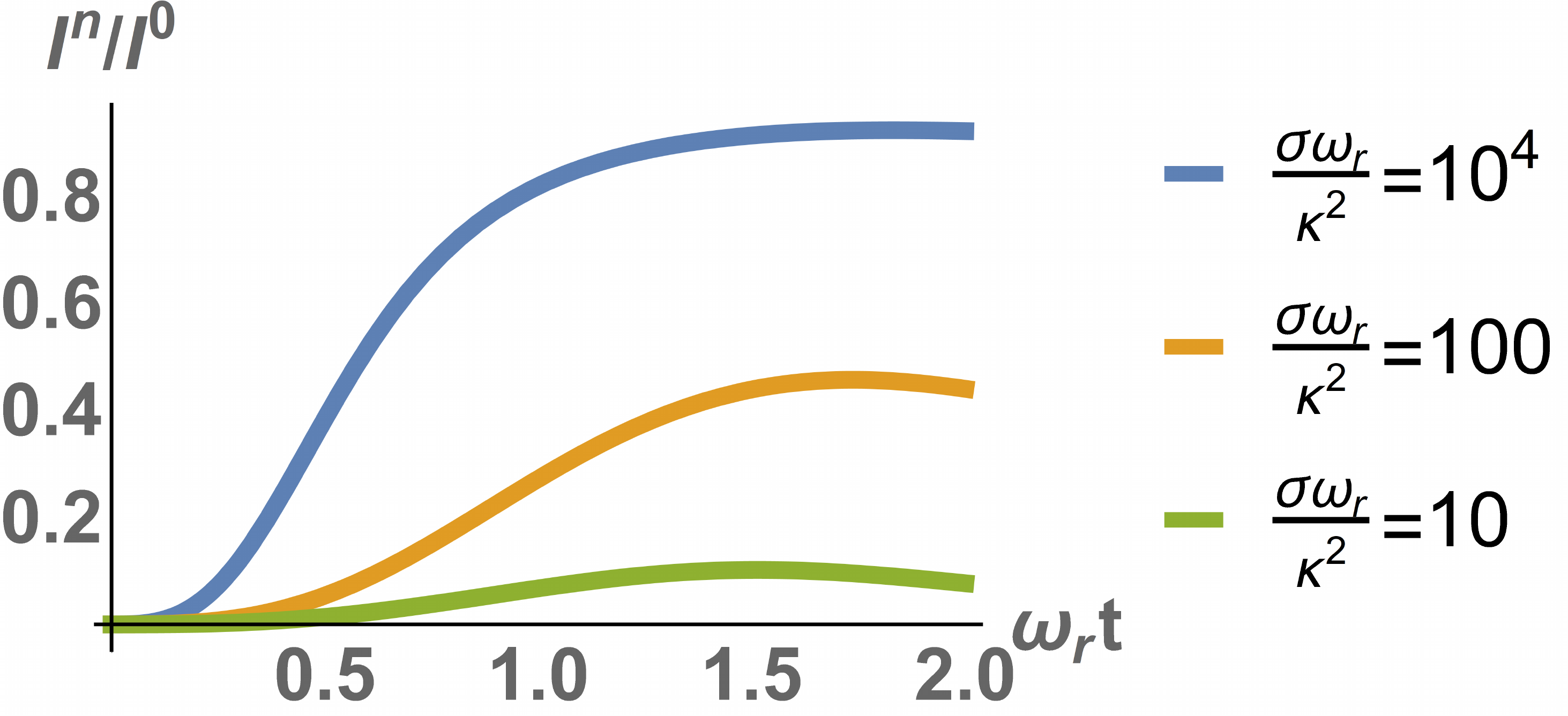}}
\end{center}
\caption{ The FI with dephasing ($I^{n}$) divided by the noiseless FI ($I^{0}$), for different values of $\frac{\sigma\omega_{ \rm{r}  }}{\kappa^{2}}.$
 Note that as $\frac{\sigma\omega_{ \rm{r}  }}{\kappa^{2}}$ gets higher the noisy FI can get closer to the noiseless FI,
 however there is a minimal time from which this can be achieved. This minimal time goes as $\sim\frac{\kappa^{1/3}}{\sigma^{2/3}\omega_{ \rm{r}  }^{2/3}}$.      }
\label{dephasing}
\end{figure}  

We therefore observe that Markovian noise on the qubit (such as dephasing) imposes a resolution limit, this invokes a natural question: can error correction protocols remove these limitations imposed by Markovian noise?
Note that this question is not analogous to the achievability of Heisenberg scaling which was addressed in \cite{sekatski2017quantum,zhou2018achieving,demkowicz2017adaptive}.
The prospects for quantum error correction in these resolution problems are left for future work, however we point out that this limit can be eliminated provided that errors can be detected, since one does not need to correct the errors. 
Given that error detection is possible, we can postselect the measurements without error and perform estimation according to them.
 These errors thus reduce the precision, but they do not impose a limitation. 
 This implies that the error correction condition to remove limitation in this problem should be different from the error correction condition for Heisenberg scaling.
 Let us denote the jump operators in the Master equation as $\left\{ L_{j}\right\} _{j}.$ For Heisenberg scaling, the error correction condition is $H\notin\text{span}\left\{I,L_{j},L_{j}^{\dagger}, L_{i}^{\dagger}L_{j}\right\} _{i,,j}$  \cite{sekatski2017quantum,zhou2018achieving,demkowicz2017adaptive},
 since we need both to detect and to correct.
In this case, since we need only to detect, the error correction condition should be: $H\notin\text{span}\left\{I, L_{j}^{\dagger},L_{j}\right\} _{j}.$
This implies, for example, that if the only noise source is amplitude damping (namely the only jump operator is $\sigma_{-}$), then detection is possible, and we can overcome the limitation.

\section{Superresolution with QFT}
\label{sec:QFT}
Consider the Hamiltonian in supplementary eq. \ref{Hamiltonian2}, and assume a relatively long coherence time (in which phases and amplitudes are constant), such that sampling is performed.
Namely Ramsey measurements are performed in different times, as it is described in \cite{schmitt2017submillihertz,boss2017quantum,glenn2018high}. The length of each measurement is $\tau$ and the total sampling time is $T=N\tau.$
The standard way to analyze this data is to perform a Fourier transform and fit the power spectrum. However this method suffers from a resolution limit \cite{rotem2017limits},
since the probability is symmetric with respect to $\omega_{ \rm{r}  }$ and the noise does not vanish. 

We claim that storing the data in a quantum state (using memory qubits) and using the same trick of nullifying the projection noise, then resolution limit can be beaten.
In a standard Ramsey experiment the state of the probe, after phase accumulation, is $\frac{1}{\sqrt{2}}\left(|0\rangle+e^{i \phi}|1\rangle\right).$
If we entangle the probe to memory qubits in each phase acquisition, the following state of the memory qubits can be generated:
\begin{equation}     
|\psi\rangle=\frac{1}{\sqrt{N}}\left(\underset{j=0}{\overset{N-1}{\sum}}|j\rangle e^{i\phi_{j}}\right),
\end{equation}
 where $\phi_{j}=\tau\left[\underset{i}{\sum}A_{i}\cos\left(\omega_{i} t_{j}  \right)+B_{i}\sin\left(\omega_{i}  t_{j}   \right)\right], \; (t_{j}=j\tau).    $ 
 The idea is that for $\omega_{ \rm{r}  }=0$ only harmonics of $\omega_{ \rm{s}  }$ can be measured, and the probability to measure the other frequencies goes as $\omega_{ \rm{r}  }^{2}.$
To see this note that
\begin{equation}
 \phi_{j}\left(\omega_{ \rm{r}  }=0\right)=\tau\left[\underset{i}{\sum}A_{i}\cos\left(\omega_{ \rm{s}  }t_{j}\right)+B_{i}\sin\left(\omega_{ \rm{s}  }t_{j}\right)\right]=\Omega\sin\left(\omega_{ \rm{s}  }t_{j}+\varphi\right),
\end{equation}
and therefore: 
\begin{equation}
e^{i\phi_{j} \left(\omega_{ \rm{r}  }=0\right)  }=\underset{k=-\infty}{\overset{\infty}{\sum}}J_{k}\left(\Omega\right)\exp\left(ik\varphi\right)\exp\left(ik\omega_{ \rm{s}  }j\tau\right),
\end{equation}
where this expansion to harmonics of $\omega_{ \rm{s}  }$ is the Jacobi-Anger expansion. 

Since we want to make sure that $\omega_{ \rm{s}  }$ (and integer multiples of it) will be included in the Fourier basis, we need to set $T=\frac{2\pi}{\omega_{ \rm{s}  }}m$ (integer $m$).
In order to avoid too many harmonics, we also set $\tau=\frac{2\pi}{\omega_{ \rm{s}  }}\frac{1}{n}$ (integer $n$).

It is now simple to see that with this choice, the only frequencies that can be measured in QFT are: $ 0,\omega_{ \rm{s}  },...,\left(n-1\right)\omega_{ \rm{s}  },$ 
as the state reads:
\begin{equation}
|\psi_{0}\rangle=\underset{l=0}{\overset{n-1}{\sum}}a_{l}|l\omega_{ \rm{s}  }\rangle,
\end{equation}
where $a_{l}=\underset{k=-\infty}{\overset{\infty}{\sum}}J_{nk+l}\left(\Omega\right)\exp\left(i\left(nk+l\right)\varphi\right).$

So for example given a noise model of a random phase, the density matrix is diagonal in the Fourier basis: $\rho=\underset{l=0}{\overset{n-1}{\sum}}p_{l}|l\omega_{ \rm{s}  }\rangle\langle l\omega_{ \rm{s}  }|,$ where $p_{l}=\underset{k=-\infty}{\overset{\infty}{\sum}}|J_{nk+l}\left(\Omega\right)|^{2},$  
hence the optimal measurement basis is the Fourier basis (and it is enough to measure whether we get harmonics of $\omega_{ \rm{s}  }$ or not).

 \begin{figure}[h]
\begin{center}
\centering
{\includegraphics[width=10 cm]{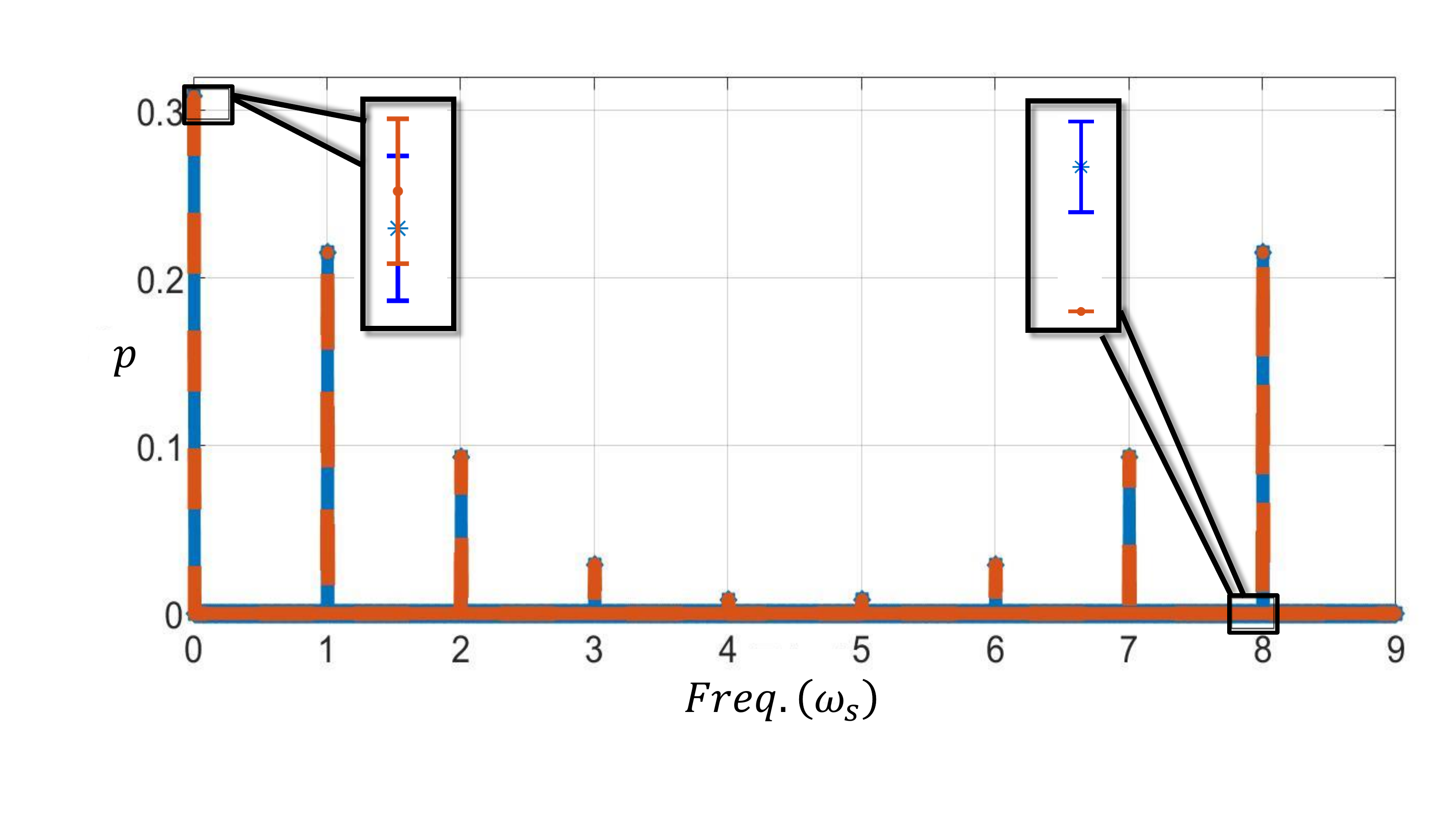}}
\end{center}
\caption{ Illustration of superresolution with QFT: probabilities of the different frequencies for $\omega_{ \rm{r}  }=0$ (dashed, orange lines) and $\omega_{ \rm{r}  }T=0.05$ (solid, blue line).
The probabilities to measure frequencies that are not harmonics of $\omega_{ \rm{s}  }$ goes as $\omega_{ \rm{r}  }^2,$ and thus superresolution can be achieved by measuring these frequencies.
In this illustration, the quadratures are random with $\sigma \tau=1,$ and the number of repetitions is $10^{5}.$  }
\label{QFT}
\end{figure}  

Let us now find the probability to measure frequencies that are not harmonics of $\omega_{ \rm{s}  }$, for $\omega_{ \rm{r}  }T \ll 1.$ 
We denote the projector on the other frequencies (not harmonics of $\omega_{ \rm{s}  }$) as $\Pi,$ so we are interested in finding $p=|\Pi|\psi\rangle|^{2}.$ 
Of course: $|\psi\rangle=|\psi_{0}\rangle+\omega_{ \rm{r}  }\frac{d|\psi\rangle}{d\omega_{ \rm{r}  }}+\mathcal{O}\left(\omega_{ \rm{r}  }^{2}\right),$ and since $\Pi|\psi_{0}\rangle=0,$ we get that: $p\approx\omega_{ \rm{r}  }^{2}|\Pi\frac{d|\psi\rangle}{d\omega_{ \rm{r}  }}|^{2}.$
Now:
\begin{equation*}
\frac{d|\psi\rangle}{d\omega_{ \rm{r}  }}=\underset{j=1}{\overset{N}{\sum}}\exp\left(i\phi_{j}\left(\omega_{ \rm{r}  }=0\right)\right)\left(i\omega_{ \rm{r}  }t_{j}\right)\left[\left(B_{1}-B_{2}\right)\tau\cos\left(\omega_{ \rm{s}  }t_{j}\right)+\left(A_{2}-A_{1}\right)\tau\sin\left(\omega_{ \rm{s}  }t_{j}\right)\right]|j\rangle,
\end{equation*}  
note that we can expand:
\begin{equation}
\exp\left(i\phi_{j}\left(\omega_{ \rm{r}  }=0\right)\right)\left[\left(B_{1}-B_{2}\right)\tau\cos\left(\omega_{ \rm{s}  }t_{j}\right)+\left(A_{2}-A_{1}\right)\tau\sin\left(\omega_{ \rm{s}  }t_{j}\right)\right]=\underset{l=0}{\overset{n-1}{\sum}}b_{l}\exp\left(il\omega_{ \rm{s}  }t_{j}\right),
\end{equation}
Therefore:
\begin{equation}
\frac{d|\psi\rangle}{d\omega_{ \rm{r}  }}=\frac{i}{\sqrt{N}}\underset{l=0}{\overset{n-1}{\sum}}b_{l}\underset{j=0}{\overset{N-1}{\sum}}t_{j}\exp\left(il\omega_{ \rm{s}  }t_{j}\right) |j\rangle.
\end{equation}
For convenience let us denote $|r_{k}\rangle=\frac{1}{\sqrt{N}}\underset{j=1}{\overset{N}{\sum}}t_{j}\exp\left(ik\omega_{ \rm{s}  }t_{j}\right)|j\rangle,$ then with this notation: 
\begin{equation}
\frac{d|\psi\rangle}{d\omega_{ \rm{r}  }}=i\underset{l=0}{\overset{n-1}{\sum}}b_{l}|r_{l}\rangle.
\end{equation}
Now given that $T\gg \frac{2 \pi}{\omega_{ \rm{s}  }},$ the state $|r_{l}\rangle$ will have a non-negligible overlap only with frequencies close enough to $|l \omega_{ \rm{s}  }\rangle,$
this leads us to make two approximations:  $i\neq k\Rightarrow\langle r_{k}|\Pi|r_{i}\rangle=0$ (different $|r_{k}\rangle$'s overlap orthogonal frequencies) and
$|\Pi|r_{k}\rangle|^{2}=\langle r_{k}|r_{k}\rangle-|\langle k\omega_{ \rm{s}  }|r_{k}\rangle|^{2}$ (the only harmonic that overlaps $|r_{k}\rangle$ is $|k\omega_{ \rm{s}  }\rangle$).

Due to the first approximation:
\begin{equation}
p\approx\omega_{ \rm{r}  }^{2}|\Pi\frac{d|\psi\rangle}{d\omega_{ \rm{r}  }}|^{2}\approx\omega_{ \rm{r}  }^{2}\underset{l=0}{\overset{n-1}{\sum}}|b_{l}|^{2}|\Pi|r_{l}\rangle|^{2}.
\end{equation}
observe now that:
\begin{eqnarray}
\begin{split}
&\langle r_{k}|r_{k}\rangle&=\frac{1}{N}\underset{j}{\sum}t_{j}^{2}\approx\frac{T^{2}}{3}\\
& \langle k\omega_{ \rm{s}  }|r_{k}\rangle&=\frac{1}{N}\underset{j}{\sum}t_{j}\approx\frac{T}{2}.
\end{split}
\end{eqnarray}
Then due to the second approximation:
\begin{equation}
p=\frac{1}{12}\omega_{ \rm{r}  }^{2}T^{2}\underset{l=0}{\overset{\left(n-1\right)}{\sum}}|b_{l}|^{2}.
\end{equation}
It is now simple to see that: $\underset{l=0}{\overset{\left(n-1\right)}{\sum}}|b_{l}|^{2}=\tau^{2}\frac{1}{2}\left[\left(A_{1}-A_{2}\right)^{2}+\left(B_{1}-B_{2}\right)^{2}\right].$
Therefore taking the model of random quadratures (each with variance $\sigma^2$), we get: $p=\frac{1}{6}\omega_{ \rm{r}  }^{2}T^{2}\left(\sigma\tau\right)^{2}.$

\section{Different noise models}
\label{Different_noise_models}
We showed in the main text that given the following effective Hamiltonian (this already takes into account the pulses, so all the relevant factors have been absorbed into the amplitudes):
\begin{equation}
H=\left[ A_{1}\cos\left(\delta_{1}t\right)+B_{1}\sin\left(\delta_{1}t\right)+A_{2}\cos\left(\delta_{2}t\right)+B_{2}\sin\left(\delta_{2}t\right)  \right] \sigma_{z},
\end{equation}
and a certain noise model of the amplitudes, the FI is calculated according to the average transition probability:
\begin{equation}
p=\int \sin^{2}\left(\phi\right) \underset{i}{\Pi} p\left(A_{i}\right)p\left(B_{i}\right)\:dA_{i}\:dB_{i}.
\end{equation}
Using the control method proposed in this paper (applying $\pi$ pulses such that $\delta_{s}t=2\pi$), we obtain that $\phi\approx\frac{\left(A_{1}-A_{2}\right)}{2 \pi}\omega_{ \rm{r}  }t^{2}\rightarrow p_{a}\propto\omega_{ \rm{r}  }^{2}.$
Therefore a non vanishing $I_{ \rm{r}  }$ is achieved as long as $\int\left(A_{1}-A_{2}\right)^{2}p\left(\text{{\bf A}},\text{{\bf B}}\right)\:d\text{{\bf A}}\:d\text{{\bf B}}\neq0.$ 
Since our primary interest is in NMR we assumed the noise model relevant to unpolarized NMR in which $A_{i},B_{i}$ are Gaussian i.i.d. with a distribution of $N\left(0,\sigma\right).$
Assuming this noise model, the average transition probability,$p,$ is given by:
\begin{equation}
p=0.5\left(1-\exp\left(- 8 \sum_{i}    \frac{\sigma^2}{\delta_i^2}  \sin^2\left(\frac{\delta_i t}{2}  \right)    \right)   \right),
\end{equation}
For $\delta_{s}t=2\pi,$ $p_{a}\approx\frac{2\sigma^{2}}{\omega_{ \rm{s}  }^{2}}\omega_{ \rm{r}  }^{2}t^{2},$ and thus $I_{ \rm{r}  }=\frac{2\sigma^{2}t^{4}}{\pi^{2}}.$ 

For classical signals (such as microwave signals generated by AC wires) a different noise model should be taken into account. For these signals the amplitude of the field ($\sqrt{A^{2}+B^{2}}$) is constant, while the phase ($\text{arctan}\left(\frac{B}{A}\right)$) distributes uniformly. It is easy to verify that in this case $\int\left(A_{1}-A_{2}\right)^{2}p\left(\text{{\bf A}},\text{{\bf B}}\right)\:d\text{{\bf A}}\:d\text{{\bf B}}\neq0,$ and thus a finite FI is achieved.
A detailed analysis shows that:
\begin{equation}  
p=\frac{1}{2}\left(1-J_{0}\left(\frac{4\Omega_{1}}{\delta_{1}}\sin\left(\frac{\delta_{1}t}{2}\right)\right)J_{0}\left(\frac{4\Omega_{2}}{\delta_{2}}\sin\left(\frac{\delta_{2}t}{2}\right)\right)\right),
\end{equation}
where $\Omega_{i}=\sqrt{A_{i}^{2}+B_{i}^{2}},$ and these amplitudes are constants and identical. Taking $\delta_{s}t=2\pi,$ we get:
\begin{equation}
p\approx\frac{\left(\Omega t\right)^{2}}{\left(2\pi\right)^{2}}\omega_{ \rm{r}  }^{2}t^{2}\Rightarrow I=\frac{4\Omega^{2}}{\left(2\pi\right)^{2}}t^{4}.
\end{equation}

\end{widetext}

\end{document}